\documentclass{aa}  

\usepackage{txfonts}
\usepackage{orcidlink}

\newcommand{\arepo}{{\sc Arepo}}
\newcommand{\sevn}{{\sc sevn}}
\newcommand{\parsec}{{\sc Parsec}}

\newcommand{\arepogw}{{\sc Arepo-GW}}
\newcommand{\SEVNbench}{\textit{SEVNbenchmark2401}}

\usepackage{hyperref}
\usepackage{comment}

\hypersetup{colorlinks=true,linkcolor=[rgb]{1.,0.2,0.2},citecolor=[rgb]{0.1,0.4,1.},filecolor=[rgb]{0.7,0.2,0.2},urlcolor=[rgb]{0.7,0.2,0.2}}

\usepackage{graphicx}

\newcommand{\rev}[1]{#1}

\defcitealias{GWTC4}{LVK25}

\begin{document}

   \title{Modeling gravitational wave sources in the MillenniumTNG simulations}

   \author{Federico Marinacci\inst{1,2}\orcidlink{0000-0003-3816-7028}
          \and
          Marco Baldi\inst{1,2,3}\orcidlink{0000-0003-4145-1943}
          \and
          Giuliano Iorio\inst{4}\orcidlink{0000-0003-0293-503X}
          \and
          M.~Celeste Artale\inst{5}\orcidlink{0000-0003-0570-785X}
          \and
          Michela Mapelli\inst{6,7}\orcidlink{0000-0001-8799-2548}
          \and
          \\
          Volker Springel\inst{8}\orcidlink{
          0000-0001-5976-4599}
          \and
          Sownak Bose\inst{9}\orcidlink{0000-0002-0974-5266}
          \and
          Lars Hernquist\inst{10}\orcidlink{0000-0001-6950-1629}
          }

   \institute{Dipartimento di Fisica e Astronomia "Augusto Righi", Universit\`a di Bologna, via Piero Gobetti 93/2, I-40129 Bologna, Italy.\\\email{federico.marinacci2@unibo.it}
   \and
             INAF, Osservatorio di Astrofisica e Scienza dello Spazio di Bologna, via Piero Gobetti 93/3, I-40129 Bologna, Italy.
             \and
             \rev{INFN, Sezione di Bologna, viale Berti Pichat 6/2, I-40127 Bologna, Italy.}
             \and
            Institut de Ciències del Cosmos (ICCUB), Universitat de Barcelona (UB), c. Martí i Franquès 1, 08028 Barcelona, Spain.
             \and
             Universidad Andres Bello, Facultad de Ciencias Exactas, Departamento de Fisica y Astronomia, Instituto de Astrofisica, Fernandez Concha 700, Las Condes, Santiago RM, Chile.
             \and
             Universit\"at Heidelberg, Zentrum f\"ur Astronomie, Institut f\"ur Theoretische Astrophysik, Albert Ueberle Str. 2, D-69120 Heidelberg, Germany.
             \and
             Universit\"at Heidelberg, Interdisziplin\"ares Zentrum f\"ur Wissenschaftliches Rechnen, D-69120 Heidelberg, Germany.
             \and
             Max-Planck-Institut f\"ur Astrophysik, Karl-Schwarzschild-Strasse 1, D-85748 Garching, Germany.
             \and
             Institute for Computational Cosmology, Department of Physics, Durham University, South Road, Durham DH13LE, UK.
             \and
             Center for Astrophysics | Harvard \& Smithsonian, 60 Garden St, Cambridge, MA 02138, USA.
             }

  \date{Received 7 October 2025 / Accepted 9 January 2026}

  \abstract{We introduce a flexible framework for building gravitational wave (GW) event catalogs in hydrodynamic simulations of galaxy formation. Our framework couples the state-of-the-art binary population synthesis code \sevn\ with \arepogw\ -- a module fully integrated into the moving-mesh code \arepo\ -- to assign merger events of binary compact objects to stellar particles in simulations by stochastically sampling merger tables generated with \sevn. \arepogw\ supports both on-the-fly operation, producing event catalogs during simulations, and post-processing, using snapshots from existing runs. The algorithm is fully parallel and can be readily adapted to outputs from simulation codes beyond \arepo. To demonstrate the capabilities of our new framework, we applied \arepogw\ in post-processing to simulations from the MillenniumTNG suite, including its flagship box -- one of the largest full-physics cosmological simulations to date. We investigate key properties of the resulting GW event catalog, built on \sevn\ predictions, focusing on comoving merger rates, formation efficiencies, delay-time distributions, and progenitor mass and metallicity distributions. 
  We also examine how these properties vary with simulated volume. 
  We find that GW progenitor rates closely track simulated star formation histories and are generally consistent with current observational constraints at low redshift, aside from an excess -- by a factor of $\sim 4.5$ -- in binary black hole mergers that is in line with trends reported in the literature. Moreover, our binary black hole merger rates decline more slowly with redshift than current observational estimates for $z \lesssim 1$. Finally, the analysis of progenitor mass functions across different formation channels reveals only mild redshift evolution, in agreement with earlier studies, while the binary black hole mass function displays features compatible with current observational determinations. These findings highlight the potential of our novel framework to enable detailed predictions for upcoming GW surveys within a full cosmological context.}


  \maketitle

\section{Introduction}

The first direct detection of gravitational waves (GWs) by LIGO in 2015 ushered in the era of GW astronomy. Since then, a growing number of detections has enabled detailed studies of binary compact objects, such as binary black holes (BBHs), binary neutron stars (BNSs), and black hole–neutron star (BHNS) systems. The fourth observing run (O4) of the LIGO-Virgo-KAGRA (LVK) collaboration has further increased the number of observed events, recently collected in the Gravitational-Wave Transient Catalog 4 \citep[GWTC-4;][hereafter \citetalias{GWTC4}]{GWTC4}. These new results significantly improve our ability to probe the population demographics of merging compact objects and their astrophysical environments \citep[see also][for previous GWTC releases]{Abbott2019_GWTC1,GWTC2,GWTC3}.
Next-generation GW detectors will expand the observable volume dramatically, allowing for exploration of BBH mergers out to very high redshifts, potentially reaching $z\lesssim100$ \citep[see e.g.,][]{CosmicExplorer2021,Branchesi2023,ET_BlueBook2025}.

To interpret the population of GW events, it is essential to establish a physically motivated framework that connects compact binary formation to the cosmic history of star formation and chemical enrichment of the Universe. Several methods have been proposed to address this challenge. A possible solution is to use semi-analytic techniques or observationally calibrated star formation histories, convolved with parameterized binary formation prescriptions and merger time distributions, to estimate merger rate densities across cosmic time \citep{Adhikari2020,Rauf2023,Vijaykumar2024,Dehghani2025}. Although computationally inexpensive, these approaches cannot often track mergers within the complex dynamical environments of evolving galaxies and halos.

A more recent class of models has attempted to overcome this issue by combining merger trees -- derived from $N$-body or hydrodynamical simulations -- or empirical galaxy scaling relations with (simplified) binary evolution models \rev{\citep{2012ApJ...759...52D,2013ApJ...779...72D,2016MNRAS.463L..31L,2019ApJ...881..157B,2019MNRAS.490.3740N,2020MNRAS.493L...6T,2021ApJ...907..110B,Santoliquido2022,2023ApJ...948..105V,2024ApJ...976...23B,2024AnP...53600170C,Sgalletta2025}}. 
This hybrid approach captures certain aspects of hierarchical galaxy growth and metallicity enrichment, but still lacks a mapping of the spatial distribution of events and relies on assumptions underlying the empirical scaling relations and binary evolution models. \rev{Similar techniques have also been applied to estimate the GW background produced by mergers of supermassive black holes at the centers of galaxies \citep[see, e.g.,][]{2017MNRAS.464.3131K, 2017MNRAS.471.4508K, 2020MNRAS.498..537S}.}

Finally, a more comprehensive (and computationally expensive) method involves the post-processing of cosmological hydrodynamical simulations with stellar evolution and binary population synthesis models. In these studies, stellar particles are assigned binary merger probabilities based on metallicity, age, and star formation history, enabling predictions for merger rates and host galaxy properties \citep{Mapelli2017,2017MNRAS.471L.105S,Mapelli2018,Artale2019,Artale2020,Artale2020b,Rose2021,Perna2022,Sgalletta2023}.

Despite their success, all these strategies face several limitations. First, simulations with small cosmological volumes (\( \lesssim 100^3 \, \mathrm{Mpc}^3 \)) often fail to sample rare, high-density environments such as galaxy clusters, which may host elevated merger rates due to their dense star-forming regions or dynamical assembly histories. Also, semi-analytical and empirical schemes are unable to capture feedback-regulated star formation and enrichment in a self-consistent manner, which may be essential for robust predictions of merger rates, delay times, and host galaxy properties.

To overcome these challenges, we advocate for an approach that couples binary population synthesis models directly into large-volume cosmological hydrodynamical simulations, enabling compact binary formation and evolution to be computed self-consistently from the simulation properties, either on-the-fly during the simulation runtime or in post-processing starting from the simulation outputs. This method ensures that the local physical conditions such as gas density, metallicity, stellar age, and dynamical environment, are accurately reflected in the properties and evolution of the resulting compact binary populations. In this framework, the cosmic evolution of GW sources is naturally coupled to galaxy formation physics, feedback, and structure growth, 
also providing a way to directly probe their spatial distribution, clustering, and bias with respect to the stellar and matter density fields.

In this work, we integrate results obtained from the binary population synthesis code \sevn\ \citep[Stellar EVolution for N-body][]{Spera2015,Spera2019,Mapelli2020,Iorio2023} directly into the large-volume hydrodynamical simulations performed with the moving-mesh code \arepo\ \citep{Springel2010, Pakmor2016}, through a fully integrated module that we refer to as \arepogw. As a first application, we employ this framework on the cosmological simulations of the MillenniumTNG (hereafter MTNG) project \citep{Pakmor2023}. These simulations span a comoving volume of \( \approx 740^3 \, \mathrm{Mpc}^3 \) with full hydrodynamics, metal-dependent cooling, stellar and AGN feedback, and a calibrated galaxy formation model consistent with key observables. By combining \sevn\ with the MTNG simulation suite, we can track the formation, evolution, and mergers of compact binaries in self-consistent connection with the local star formation rate, gas-phase metallicity, and galactic environment of each stellar  particle.

This work presents the methodology of  -- and initial results obtained with -- our new framework. Specifically, we focus on predictions for the redshift evolution of GW merger rates, the metallicity distribution of progenitors, and the dependence of merger features on simulated volume and the global properties of the MTNG flagship simulation. Our goal is to provide a physically motivated model to interpret current and future GW observations, while also enabling cross-correlation with electromagnetic surveys and cosmological tracers. Ultimately, our approach opens new avenues for using GWs not only as probes of stellar evolution, but also as tools to investigate the structure and evolution of the Universe.

The paper is organized as follows. In Section \ref{sec:MTNG}, we provide a brief overview of the MTNG project, including a description of the key physical processes implemented in its galaxy formation model and the properties of the simulation boxes analyzed in this work. 
Section \ref{sec:method} outlines the methodology developed to generate GW event catalogues by associating stellar binary mergers with star particles in hydrodynamical galaxy formation simulations. In Section \ref{sec:results},
we present the main results of our analysis, focusing on the key properties of GW progenitor events in the flagship simulation box of the MTNG project. Finally, in Section \ref{sec:conclusion}, we summarize our findings and conclude. \rev{Appendix~\ref{sec:appendixA} presents a quantitative comparison between the two operational modes of the \arepogw\ framework while Appendix \ref{sec:appendixB}} extends the analysis of GW progenitor properties to all the other simulation boxes from the MTNG suite included in this study.

\section{The MillenniumTNG project}\label{sec:MTNG}

The MTNG project is a suite of cosmological simulations that merge the large simulated volume of the Millennium simulation \citep{Millennium} with a state-of-the-art galaxy formation physics model. 
Therefore, MTNG represents a unique combination between the need for extremely large simulation volumes for cosmological applications and recent advancements in hydrodynamical simulations that enable the self-consistent generation of a realistic galaxy population.

The galaxy formation physics model of MTNG is based on that of the highly successful IllustrisTNG simulations and includes as core processes:
(i) an explicit treatment of the interstellar medium  based on a physically motivated effective equation of state \citep{Springel2003},  (ii) a stochastic prescription of star formation \citep{Springel2003}, (iii) primordial and metal line cooling \citep{Vogelsberger2013, Pillepich2018}, (iv) mass return from stars and metal enrichment of the interstellar gas by core-collapse supernovae, type Ia supernovae, and asymptotic giant branch (AGB) stars \citep{Vogelsberger2013}, (v) an effective model for  scaling the properties (speed, mass, and metal loading) of galactic winds \citep{Pillepich2018},
and (vi) a model for the seeding, growth and feedback from supermassive black holes \citep{Weinberger2017}. We refer the interested reader to \citet{Weinberger2017} and \citet{Pillepich2018} for further details about the physical processes implemented in IllustrisTNG.

Here, we analyze the hydrodynamical full physics boxes of the MTNG project, which are evolved with the moving-mesh code \arepo\ \citep{Springel2010, Pakmor2016}. We will focus in particular on the flagship run MTNG740, which has been presented in detail in \citet{Pakmor2023}. This realization of the MTNG simulations features a cubic periodic box with side length 500 $h^{-1}$Mpc ($\approx 740$ Mpc). The initial conditions are generated at $z=63$, and are sampled with
$4320^3$ dark matter particles of mass $1.7 \times 10^8 \,{\rm M_\odot}$ and $4320^3$ gas cells with an initial mass of $3.1 \times 10^7 \,{\rm M_\odot}$, which is the target baryonic mass resolution\footnote{\arepo\ implements a refinement/derefinement scheme that keeps the mass of a gas cell within a factor of two from this reference value ($m_{\rm target}$). Cells with mass $m > 2 \times m_{\rm target}$ are split into two separate cells, whereas cells with mass $m < 0.5 \times m_{\rm target}$ are dissolved into their neighbors. As star particles are generated from gas cells, their initial mass distribution will also be in this range.}. The gravitational softening length of collisionless particles (dark matter and stars) is set to $\epsilon_{\rm DM,*} = 3.7$ kpc, whereas the minimum (comoving) gravitational softening length for gas cells is set to $\epsilon_{\rm gas,min} = 370$ pc. The gravitational softening length of gas cells is adaptive and it is scaled according to their size.

We will also examine three additional full physics boxes to assess the influence of the simulated volume on our results. We indicate these additional runs with MTNG46, MTNG92, and MTNG185, respectively. They differ from the flagship simulation in terms of the simulated volume only, which is a periodic cubic box with a side length of 31.25 $h^{-1}$Mpc ($\approx 46$ Mpc) for the MTNG46 simulation, 62.5 $h^{-1}$Mpc ($\approx 92$ Mpc) for the MTNG92 realization, and 125 $h^{-1}$Mpc ($\approx 185$ Mpc) for the MTNG185 run. To maintain the same mass resolution as the flagship simulation, the number of particles is adjusted accordingly. Specifically, MTNG46 uses $2 \times 270^3$ dark matter particles and gas cells, MTNG92 employs $2 \times 540^3$ total particles, while MTNG185 contains $2 \times 1080^3$ resolution elements overall. This ensures that each dark matter particle and gas cell retains the same initial mass as in the flagship simulation. All other parameters are left unchanged. 

Finally, we note that MTNG employs the cosmological parameters from \citet{Planck2016}. Specifically, the adopted values are $\sigma_8 = 0.8159$, $n = 0.9667$, $h = 0.6774$, $\Omega_m = 0.3089$, $\Omega_b = 0.0486$, and $\Omega_\Lambda = 0.6911$. This choice was made to facilitate the comparison of the simulation outcomes with those obtained in the IllustrisTNG suite, which is based on the same cosmology. Moreover, since our approach to modeling the GW signal in MTNG can be seamlessly applied to the original IllustrisTNG simulations (see  Sec.~\ref{sec:GWsampl}), this also allows us to further assess the robustness of our results while factoring out the influence of cosmological parameter choice.

\section{Methods}\label{sec:method}

\subsection{\sevn: performing state-of-the-art binary population synthesis simulations
}\label{sec:SEVN}

To model the GW signal in MTNG, we first require a catalogue of compact binary mergers originating from stellar populations.
To build this catalogue, we use the archival dataset \SEVNbench\footnote{The dataset is publicly availabe on Zenodo at the following link: \href{https://doi.org/10.5281/zenodo.16587145}{https://doi.org/10.5281/zenodo.16587145}.} 
 that was produced with the state-of-the-art binary population synthesis code \sevn\footnote{Our analysis employed \sevn\ version 2.7.5 (commit \href{https://gitlab.com/sevncodes/sevn/-/commit/7bb74a899cde659b75284aceb3202bc7cc5a6158}{7bb74a89}). More information on \sevn\ can be found at \href{https://sevncodes.gitlab.io/sevn/}{https://sevncodes.gitlab.io/sevn/}.} \citep{Spera2015,Spera2019, Mapelli2020, Iorio2023}.
The dataset contains the evolution of $2 \times 10^7$ stellar binary systems for each of 15 representative metallicities in the range $0.0001 \leq Z \leq 0.03$\footnote{$Z$ denotes the mass fraction of elements heavier than helium. The specific values considered are $Z = 0.0001$, 0.0002, 0.0004, 0.0006, 0.0008, 0.001, 0.002, 0.004, 0.006, 0.008, 0.01, 0.014, 0.017, 0.02, and 0.03.}, for a total of $3 \times 10^8$ binaries.

\sevn\ models the stellar evolution by interpolating (in mass and metallicity) sets of precomputed stellar tracks for H-rich stars and pure-He stars (i.e. stars stripped by their H-rich envelope).
Further information on the stellar models and interpolation schemes implemented in \sevn\ is provided in \citet{Iorio2023} and \citet{Korb2025}.
To produce the \SEVNbench\ dataset, the updated \parsec\ stellar evolution models \citep{Bressan2012, Chen2015, Costa2019, Nguyen2022} as presented in \citet{Costa2025}, were adopted among the available stellar tracks.

\sevn\ assumes that all stars with a final carbon–oxygen core mass exceeding $1.38 \ \mathrm{M}_\odot$ undergo core collapse, forming a compact remnant, either a black hole or a neutron star.
To model the core-collapse phase, the DelayedGauNS prescription \citep[see Appendix A2 in][]{Iorio2023} is used.
This prescription is based on the delayed supernova model by \citet{Fryer2012} and produces no mass gap between neutron stars and black holes, in agreement with recent observational evidence \citep[e.g.][]{GW190814,Wyrzykowski2020,Ray2025}.
Compact remnants with masses above $3 \ \mathrm{M}_\odot$ are classified as black holes, while those below this threshold are considered neutron stars. 
Black hole masses are further reduced to account for neutrino-driven mass loss, applying the correction proposed by \citet[][equation 13 in \citealt{Iorio2023}]{Zevin2020}.
The \citet{Fryer2012} models are known to produce neutron star masses that do not reproduce the observed distribution in the Milky Way \citep[see e.g.][]{Giacobbo2018, Kruckow2018, VGomez2018}. Therefore, to generate the dataset, neutron star masses are re-sampled from a Gaussian distribution with a mean of 1.33 $\mathrm{M}_\odot$, a standard deviation of 0.09 $\mathrm{M}_\odot$, and lower and upper limits of 1.1 $\mathrm{M}_\odot$ and 3 $\mathrm{M}_\odot$, respectively, so that their  mass distribution is consistent with Galactic observations \citep{Ozel2016}.

Toward the end of their evolution, massive stars can experience hydrodynamical instabilities triggered by the efficient production of electron-positron pairs.
This process can lead to either pulsational pair-instability, characterized by repeated mass-loss episodes \citep{Yoshida2016,Woosley2017}, or to pair-instability supernovae, in which a single, extremely energetic pulse completely disrupts the star, leaving no remnant \citep{Heger2003,Costa2021}.
In \sevn\ these effects are accounted for by the fiducial {\it M20} model \citep{Mapelli2020, Iorio2023}, which is based on hydrodynamical simulations of \citet{Woosley2017}.
In this model, the fate of a massive star is determined by the mass of its helium core ($m_\mathrm{He}$) at the onset of carbon burning.
Stars with $32 \ \mathrm{M}_\odot < m_\mathrm{He} < 64 \ \mathrm{M}_\odot$ undergo pulsational pair-instability, while those with $64 \ \mathrm{M}_\odot < m_\mathrm{He} < 135 \ \mathrm{M}_\odot$ explode as pair-instability supernovae, leaving no remnant.
Stars with even more massive helium cores collapse directly into black holes.

At the formation of a black hole or neutron star, the remnant receives a natal kick. These kicks are modeled following  \citet{Giacobbo2020}. Specifically, the kick magnitude is drawn from a Maxwellian distribution with a root-mean-square velocity of $265 \ \mathrm{km\,s^{-1}}$ \citep{Hobbs2005}, and then rescaled by a factor proportional to $m_\mathrm{ej}/m_\mathrm{rem}$, with $m_\mathrm{ej}$ indicating the mass of the ejecta and $m_\mathrm{rem}$ the remnant mass. The kick direction is randomly sampled assuming spherical symmetry. In the case of direct collapse ($m_\mathrm{ej} = 0$) the resulting kick is zero.
Within this framework, massive black holes and remnants formed from stripped-envelope stars typically receive smaller natal kicks. As a consequence of the kick, \sevn\  updates the orbital properties of the binary, including the possibility that the system becomes unbound.
If the binary is disrupted, \sevn\ continues to evolve the two stars as effectively single objects.

\sevn\ includes all the relevant binary processes (mass transfer, common envelope, stellar tides, gravitational wave decay) by implementing analytic and semi-analytic formalism to model the change on stellar (mass, radius, rotation) and binary (separation, eccentricy) properties.

Mass transfer processes in \sevn\ include both accretion from stellar winds and Roche-lobe (RL) overflow. Wind mass transfer is always active, while RL overflow mass transfer is triggered when one of the two stars (the donor) expands beyond its RL \citep{Eggleton1983}, transferring mass to its companion (the accretor).
The adopted \sevn\ setup  assumes that, during RL overflow mass transfer, 50\% of the mass lost by the donor is accreted by the accretor.
If the accretor is a compact object (i.e., a black hole or neutron star), accretion, whether from stellar winds or RL overflow, is limited by the Eddington accretion rate (see equation 17 in \citealt{Iorio2023}).
Any mass lost by the donor that is not accreted leaves the system as a fast, isotropic wind from the donor (in case of the winds) or from the accretor (in case of RL overflow), carrying away angular momentum from the binary.

Certain binary configurations (short separations, high mass-ratios, donor with deep convective envelopes) can lead to mass transfer that is unstable on dynamical timescales.
In \sevn, the stability of RL overflow mass transfer is assessed by comparing the donor-to-accretor mass ratio to a critical value, $q_\mathrm{crit}$ \citep{Webbink1985}. 
Stars with deep convective envelopes, such as evolved giants, are more prone to unstable mass transfer and thus have lower $q_\mathrm{crit}$ values (typically $\lesssim 1$). In contrast, stars with radiative envelopes tend to allow stable mass transfer, corresponding to higher $q_\mathrm{crit}$ values (up to $\gtrsim 3$). 

The value of $q_\mathrm{crit}$ is estimated via the \sevn\ \texttt{QCRS} default prescription \citep[see][]{Iorio2023}. This is equivalent to the $q_\mathrm{crit}$ implementation in \citet{Hurley2002}, with the modification that main sequence and Hertzsprung gap donors are always assumed to undergo stable mass transfer.
If the mass transfer is unstable, the binary enters a common envelope (CE) phase \citep{Ivanova2013}.
The CE phase is treated using the energy formalism \citep{Webbink1984}, in which the outcome, envelope ejection and orbital shriking or  merger, is determined by comparing the orbital energy released during the inspiral to the binding energy of the donor’s envelope. A CE efficiency parameter of $\alpha = 1$ is employed, corresponding to the assumption that 100\% of the available orbital energy is used to unbind the envelope. The envelope binding energy is computed using the prescription in \citet{Claeys2014}.

\sevn\ models stellar tides using the weak friction approximation developed by \citet{Hut1981}, as implemented in \citet{Hurley2002}.
Tidal effects are active throughout all phases of the simulation.
If tides are not strong enough to circularize the orbit by the onset of RL overflow, circularization at periastron is enforced by setting the new orbital semimajor axis equal to the periastron distance.

To produce the \SEVNbench\ dataset, the initial properties of the binary population are selected in the same way as in \citet[][see their Appendix A.2]{Korb2025}.
Primary masses are drawn from a \citet{Kroupa2001} initial mass function ($\propto m^{-2.35}$) in the range $5 < m_1/\mathrm{M}_\odot < 150$. 
Secondary masses are assigned using a mass ratio $q = m_2/m_1$ drawn from a power-law distribution ($\propto q^{-0.1}$), with a minimum secondary mass of $2.2 \ \mathrm{M}_\odot$. 
Orbital periods are sampled from a distribution $\propto (\log P)^{-0.55}$ over the range 1.4 days to 866 years, and eccentricities from a distribution $\propto e^{-0.42}$ in the interval [0, 0.9], with a correction for the eccentricity–period relation as described in \citet[][their eq.~3]{Moe2017}. 

This setup efficiently samples the rare population of compact-object binaries hosting black holes and neutron stars \citep[see e.g.][]{Iorio2023, Sgalletta2023, Sgalletta2025, Korb2025}.
All binary systems are evolved until both stars reach the end of their evolution, either forming compact remnants (white dwarfs, neutron stars, or black holes), or leaving no compact remnants due to stellar (e.g., pair-instability supernovae) or binary (e.g., mergers) processes. 
We refer the reader to \citet{Iorio2023} for a comprehensive description of \sevn, including alternative prescriptions and their impact on the population of binary compact objects. 

From the \SEVNbench\ dataset, we extract all binary compact objects that merge within 14 Gyr and we store the relevant properties required to model GW events in MTNG (see Sec.~\ref{sec:GWsampl}). These include: a unique identifier for each merger event, the time of merger, measured from the birth of the binary system, the individual masses of the merging objects, and the mass of the merger remnant.
Given that the MTNG simulations adopt a \citet{Chabrier2003} initial mass function with a maximum stellar mass of $M_\mathrm{max} = 100 \ \mathrm{M}_\odot$, we exclude from the selection all systems whose progenitors include a primary star exceeding this mass limit. 
For each metallicity, we produce three separate tables, corresponding to the three GW progenitor channels considered in our analysis: BBH, BHNS, and BNS mergers. The events in each table are sorted in ascending order of merger time, to facilitate the incorporation of compact binary mergers into the MTNG cosmological simulations, as described below.

\subsection{\arepogw: building GW catalogues in galaxy formation simulations}\label{sec:GWsampl}

To predict the distribution of GW events from the evolution of stellar populations modeled within full-physics galaxy formation simulations, we have developed \arepogw, a flexible and parallel module that is fully integrated into the \arepo\ code. Our module generates the distribution of GW events by stochastically sampling mergers of compact binary systems -- that can be thought of as progenitors of GW emission -- occurring within a simple stellar population (SSP). This sampling relies on the merger tables produced by the \sevn\ code (described in Sec.~\ref{sec:SEVN}), and follows an approach similar to that adopted in previous studies \citep[see, e.g.,][]{Mapelli2018,Artale2019}.

Specifically, \arepogw\ builds upon the stellar evolution infrastructure already implemented in the galaxy formation physics model of the MTNG suite. In this framework, each star particle in a simulation is treated as a SSP following a \citet{Chabrier2003} initial mass function, and is characterized by its age and metallicity.  Based on the metallicity value of a given star particle, we select the table generated from \sevn\ simulations that most closely approximates its metal content. Once a table is selected, we evaluate --  for each one of the three GW progenitor channels -- the probability $p^*$ that a merger event occurs as
\begin{equation}
p^* = \frac{M_* \,f_{\rm bin}\, f_{\rm IMF}}{M_{\rm sim}},
\label{eq:prob}
\end{equation}
where $M_{\rm sim} \approx 4.1\times 10^8$ $\mathrm{M}_\odot$ is the total mass of simulated binary systems at the selected metallicity and for the considered GW progenitor channel (excluding systems with stars more massive than 100 $\mathrm{M}_\odot$, see Sec.~\ref{sec:SEVN}); $M_*$ is the birth mass of the star particle; $f_{\rm bin} = 0.628$ is the expected binary fraction \citep{Moe2017}\footnote{The binary fraction in \cite{Moe2017} depends on the mass of the primary. The reported value is obtained by considering the whole initial mass function.}; and $f_{\rm IMF} = 0.243$ 
is a correction factor accounting for the incomplete sampling of the initial mass function due to the lower mass limits applied to primary and secondary stars in the sampled binaries (5  $\mathrm{M}_\odot$ and 2.2 $\mathrm{M}_\odot$, respectively)\footnote{To estimate the correction factor, we assume a \citet{Chabrier2003} initial mass function ranging from 0.01 to 100 $\mathrm{M}_\odot$, and a mass ratio distribution with $q$ between 0.1 and 1.}.
The first recorded event is then considered. If the current age of the stellar particle is greater than the recorded merger time -- or in other words if enough time has passed since the birth of that stellar population to have a potential binary merger -- we extract a uniformly-distributed random number $p \in [0, 1)$. If $p < p^*$, we store key binary merger event properties, namely its position within the simulation domain, time at which the event occurred, unique identifier (so that other properties of the merger can be retrieved from the \sevn\ tables for further analysis), merger type (BBH, BHNS, BNS), and the unique identifier of the star particle in the simulations to which the merger is associated. Otherwise, we move on to the next record in the table. We stop processing the star particle when its age is less than the merger time of the record being examined. This procedure is repeated for each star particle in the simulation and iterated over all time steps. For each star particle, merger events that have already been examined in previous time steps are excluded from the list of potential events.

We implement two versions of  \arepogw. The first, which has been described above, works on-the-fly while a simulation is progressing. 
The other version can be used to postprocess a snapshot of an existing large-scale simulation and it becomes particularly useful when dealing with computationally expensive simulations producing a large data volume that cannot be easily rerun to include the GW signal. This is exactly the case that we are exploring in this work. In the post-processing mode, \arepogw\ evolves each star particle of the simulation on a timestep of a predetermined length (that can be selected by the user) from its birth to the time at which the snapshot is produced. Therefore, to have a complete catalogue of GW progenitors event, it is in principle sufficient to analyse the snapshot produced by the simulation at the lowest available redshift.

To allow for the analysis of very large snapshots on relatively small machines, we design the post-processing mode of \arepogw\ in such a way that a snapshot can be processed one or several chunks at a time. In this way, we were able to produce the GW catalogue of the flagship MTNG740 simulation analysed in this work starting from its redshift-zero snapshot ($\approx 7$ TB of data volume) on our local HPC machine. Moreover, for such large-scale applications we adjust the output strategy to minimize the total volume of saved data. First, we process star particles in the same order as they appear in the snapshot, allowing the storage of their unique identifiers to be optionally disabled. For each individual star particle we then save the associated merger events contiguously and record their total number. In this way we are able to uniquely associate each binary merger event to its parent star particle in the simulation. Second, we do not store the binary merger positions as they will be identical for all the merger events generated by the same star particle. \rev{However,} the position of the parent star particle can \rev{still} be retrieved at the time (or redshift) of a specific merger event \rev{via snapshot interpolation (if enough snapshots exist) or if light-cone like outputs are available}. Despite these optimizations leading to significant storage saving, the total volume of the GW catalogue produced for the MTNG740 simulation still amounts to $\approx 30$ TB.

\rev{Both operational modes of \arepogw\ introduce only minimal computational overhead because the module is fully embedded within the stellar evolution routines already implemented in \arepo\ \citep[see][for details]{Vogelsberger2013}. The generation of GW events requires only a table lookup and a single random draw per event, along with a lightweight output buffer. As a result, any cosmological simulation that can be run with \arepo\ can also be run with the on-the-fly mode of \arepogw. The primary cost difference -- in addition to the time requested to carry out a specific simulation when the algorithm is used on-the-fly -- between the two modes is the storage requested per event: the on-the-fly mode uses $30$ bytes to store a single event compared to $8$ bytes needed in post-processing. Since the number of GW progenitor events scales directly with the total stellar mass formed in the simulation (see eq.~\ref{eq:prob}), it is largely the simulation volume (and to a lesser extent its resolution) that determines the size of the resulting GW catalogue. However, the larger storage footprint of the on-the-fly mode rarely poses a practical limitation to the applicability of the method. For example, using MTNG-like physics and resolution would produce an event catalogue of an extimated size of $\approx 30$ TB for a simulation with a volume of $(480\,{\rm Mpc})^3$ run with \arepogw\ on-the-fly. Simulations of this size are state-of-the-art and comparable to the MTNG740 flagship box analysed in this work. The on-the-fly mode has also the advantage that precise spatial positions of GW-progenitor events are recorded at formation, eliminating the need for snapshot or light-cone interpolation. This makes the on-the-fly approach particularly well suited for studies investigating the impact of different galaxy formation physics models, where rerunning the underlying hydrodynamical simulations is required in any case.}

To first order, the only difference between the post-processing and on-the-fly modes of \arepogw\ is that, \rev{as previously discussed}, in the \rev{former} it is not possible to recover the exact positions of merger events associated with the same star particle. However, upon closer inspection, one cannot expect an identical sequence of merger events for a given star particle in the two modes, even when analyzing the same simulation. This discrepancy is caused by two main effects.
First, the time step over which a star particle evolves is fixed in post-processing, but adaptive in the on-the-fly mode. Although both approaches rely on the same power-of-two time-step hierarchy \citep[see][for more details]{Springel2010}, this can still introduce small variations in the recorded merger times. 
Second, our modeling is inherently stochastic: \sevn\ tables are sampled with a probability $p^*$, defined in eq.~(\ref{eq:prob}), which is then compared to a uniformly distributed random number. Both operational modes of \arepogw\ parallelize execution through the Message Passing Interface (MPI) library and random numbers are generated independently by each MPI task. Consequently, the precise sequence of merger events also depends on the MPI task processing the parent star particle. In the on-the-fly mode, this picture becomes more complex because (star) particles can be exchanged among MPI tasks during runtime to maintain optimal workload balance in \arepo\ \citep[][]{Springel2010}. In practice, however, provided that merger events are sufficiently well sampled, stochasticity should only introduce negligible differences between the two modes. \rev{We present a more quantitative assessment of how the two operational modes of \arepogw\ affect the properties of GW progenitor events within the same simulation in Appendix~\ref{sec:appendixA}.}

We would like to stress that the approach to generate GW catalogues described in this section has a wider scope of applicability than the MTNG simulation suite. Indeed, although the algorithm we implemented depends on the stellar evolution routines adopted in the MTNG simulations, such routines are at the foundations of several galaxy formation physics models implemented in \arepo\ that are targeted to study different aspects of structure formation. These include, among others, the Illustris and IllustrisTNG models \citep[][]{Vogelsberger2014a, Pillepich2018}, on which the MTNG simulations are based and that focus on formation and evolution of a complete galaxy population, the Auriga project \citep[][]{Grand2017}, which is centered on the formation of galaxies like the Milky Way, and the SMUGGLE model  that includes more realistic prescriptions for ISM structure and stellar feedback \citep[][]{Marinacci2019}. Moreover, our approach can be easily ported to other hydrodynamical simulation codes or used to postprocess calculations not performed with \arepo\ given the limited amount of information on star particles (e.g., initial mass, birth time, metallicity, and assumed IMF) necessary as an input.

\begin{figure}
    \centering
    \includegraphics[width=0.485\textwidth]{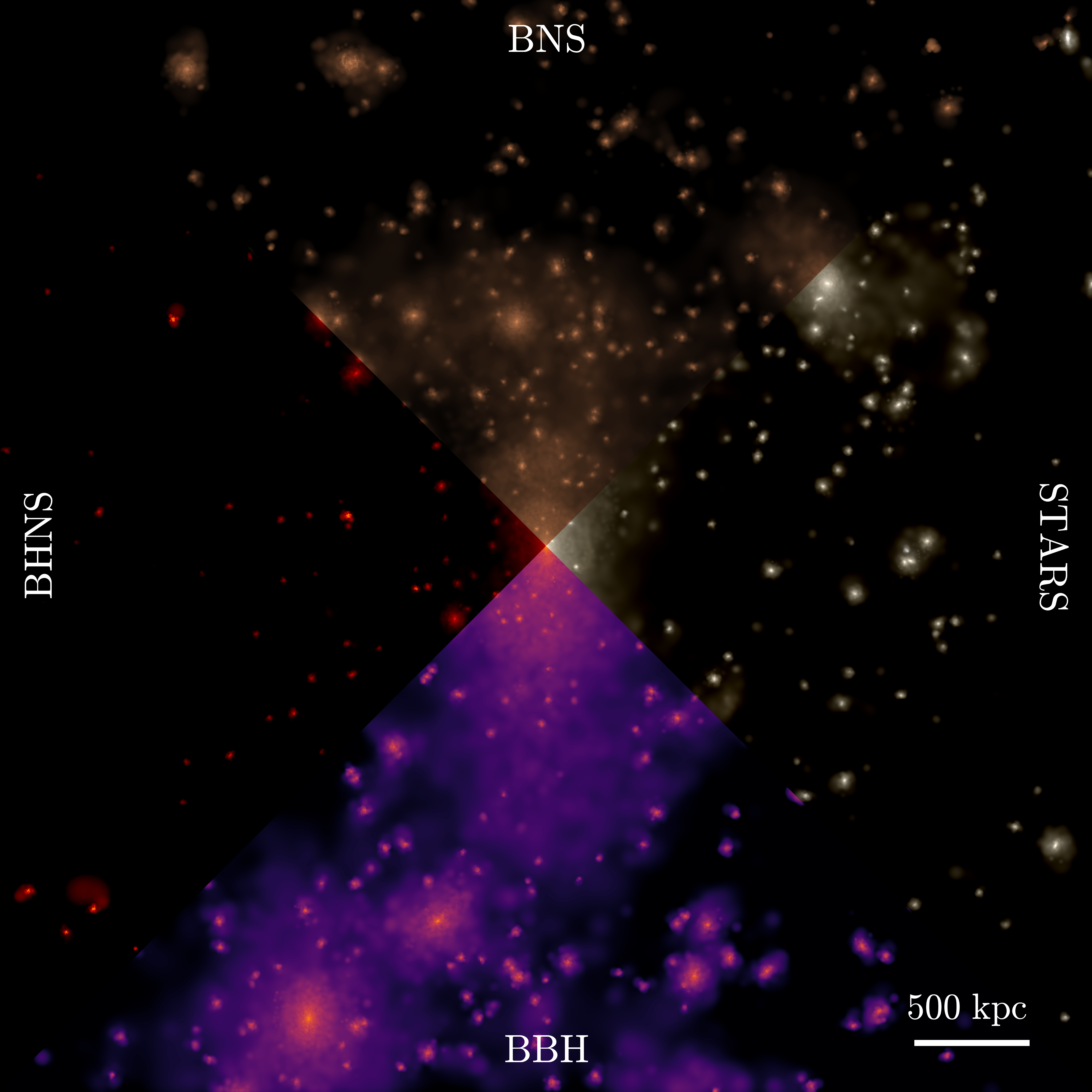}
    \caption{Map\rev{s} of projected stellar surface brightness and GW progenitors merger surface density events for the BBH, BHNS, and BNS channels (from right in a clockwise direction, as indicated in the figure). The maps are centered on the most massive FoF halo of the MTNG740 run and have a side length of 5 Mpc. The stellar map has been obtained by mapping the $g$, $r$ and $z$ photometric bands to the blue, green and red color channels of the figure, respectively, with intensities linearly scaled over the interval $[20, 30]\,{\rm mag\,arcsec^{-2}}$. For GW progenitors merger rate events only compact binary mergers occurring at $z < 0.1$ are considered. Values are logarithmically scaled over the interval $[10^{-8},10^{-2}]\,{\rm pc}^{-2}$.} 
    \label{fig1}
\end{figure}

\section{Results}\label{sec:results}

\subsection{The spatial distribution of GW progenitors events}
To visually demonstrate the capabilities of the \arepogw\ post-processing module, Fig.~\ref{fig1} presents maps of projected stellar surface brightness and the surface density of GW progenitor mergers for the three different channels -- BBH, BHNS, and BNS -- shown clockwise starting from the right. The maps are centered on the most massive friends-of-friends (FoF) structure in the MTNG740 volume and span a side length of 5 Mpc.
The stellar map is a $grz$ composite image, with photometry derived from the ages and metallicities of star particles using the spectral evolution model of \citet{2003MNRAS.344.1000B}. The maps of GW progenitor merger rates include only compact binary mergers occurring at redshifts $z < 0.1$ and are scaled to the same range of surface density values to highlight relative differences between the channels.

It is apparent that the rates of GW progenitors are closely linked to the stellar distribution of the forming structures in the simulation. Specifically, the merger events tend to cluster around regions of high stellar surface brightness, which correspond to substructures -- namely, individual galaxies. We note that the structure shown in Fig.~\ref{fig1} is representative of a galaxy cluster with a virial mass\footnote{We define the virial mass of a halo as the total mass enclosed within a radius where the mean density is 200 times the critical density of the Universe.} of $M_{\rm 200,c} \approx 8.5 \times 10^{14}\, {\rm M_\odot}$. However, a noticeable difference emerges in the spatial distribution of the GW progenitor surface densities across the different channels. BBH and BNS events feature more diffuse distributions, while BHNS mergers are more concentrated near peaks in stellar surface brightness. This behavior reflects the lower rates of BHNS mergers at low redshift in our model, which are a factor of $\sim$2–4 lower than the corresponding BBH and BNS rates (see Fig.~\ref{fig3}).
Although the $z < 0.1$ merger rates of BBH and BNS are within a factor of $\simeq 2$, the BBH merger surface density appears more spatially extended than that of BNS, despite the higher overall occurrence of BNS events. This suggests a possible difference in the delay time distributions of their progenitors, which may influence the spatial distribution of merger sites. Further details on distribution of delay times of the different GW progenitors are provided in Fig.~\ref{fig5}.

\subsection{Evolution of GW event rates}
An important quantity that can be derived from the modeling of GW sources in cosmological simulations is the time evolution of the comoving merger rate of compact binary systems across different channels. Given the approach adopted here, we expect the evolution of the comoving merger rate to be closely tied to the evolution of the cosmic star formation rate density (SFRD). Specifically, the comoving merger rate for a given channel can be expressed as:
\begin{equation}
R_{\rm ch}(t) = \int_{0}^{t} {\rm SFRD}(t')\,\Theta_{\rm ch}(t - t') \, dt',
\label{eq:rates}
\end{equation}
where $R_{\rm ch}(t)$ is the comoving merger rate at cosmic  time $t$ for a specific progenitor channel (BBH, BHNS, BNS), ${\rm SFRD}(t')$ is the comoving star formation rate density at $t'$, and $\Theta_{\rm ch}(t - t')$ is related to both the merger efficiency of the progenitors  per unit of stellar mass formed (Fig.~\ref{fig4}) and their delay time distribution\footnote{In eq.~(\ref{eq:rates}) the delay time is given by $t - t'$, where $t$ indicates the time at which a compact binary merger occurs and $t'$ is the formation time of the associated stellar population.} (Fig.~\ref{fig5}).
In our modeling, the SFRD is directly extracted from the hydrodynamical simulation, while $\Theta_{\rm ch}$ is encoded in the compact binary merger tables described in Sec.~\ref{sec:SEVN}.

  \begin{figure}
        \centering   \includegraphics[width=0.5\textwidth]{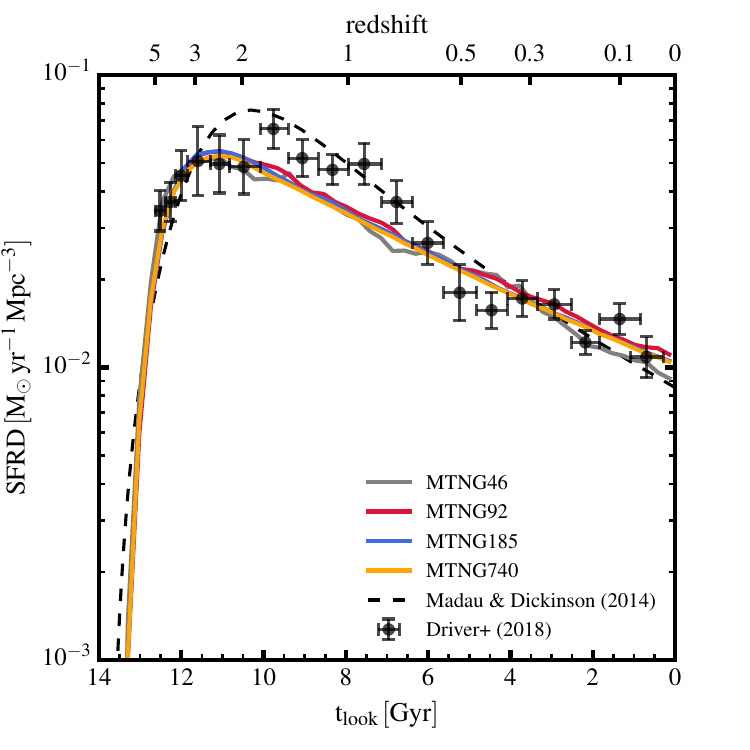}
      \caption{Comoving SFRD as a function of time/redshift in four different boxes of the MTNG simulations, as indicated in the legend. The determinations of the cosmic SFRD as reported in \citet{2014ARA&A..52..415M} and \citet{2018MNRAS.475.2891D} are also shown for comparison. Owing to the constant mass resolution of the analysed MTNG boxes, the SFRD is well-converged, with negligble differences among the runs. The MTNG SFRD peaks at earlier redshifts ($z \sim 3$ rather than $z \approx 2$) and shows a less steep decline towards lower redshift compared to the results presented in \citet{2014ARA&A..52..415M}. Overall, the SFRD values in MTNG are broadly consistent with the determinations of \citet{2014ARA&A..52..415M} and \citet{2018MNRAS.475.2891D}, although they tend to be lower by up to a factor of $\lesssim 2$ in the redshift range $z \sim 1$–$2$.
     }
         \label{fig2}
   \end{figure}

Figure~\ref{fig2} shows the SFRD obtained from the MTNG suite (solid colored lines), compared with two widely used determinations: \citet[][dashed line]{2014ARA&A..52..415M} and \citet[][points with error bars]{2018MNRAS.475.2891D}. Both SFRD determinations have been rescaled to the value of the Hubble parameter adopted in the MTNG cosmology, by multiplying them by the factors $(h/0.7)^3$ and $(h/0.7)$, respectively. The \citet{2014ARA&A..52..415M} measurements are further corrected for the different choice of the IMF (\citealt{1955ApJ...121..161S} vs. \citealt{Chabrier2003}), applying a conversion factor of $0.63$ \citep[see Table 1 in][]{2013MNRAS.430.2622D}. The SFRD of the MTNG740 run is estimated by binning stellar particles according to their ages into 50 bins in the range $0-14$ Gyr and summing for each bin their initial mass\footnote{In the MTNG galaxy formation framework star particles lose mass as a consequence of stellar evolution processes.}. The total stellar mass of each bin is then divided by the bin width ($0.28$ Gyr) and the comoving volume of the box to get the SFRD value.

The most notable feature of the SFRD derived from different MTNG realizations is its independence from the simulated volume. In other words, the SFRD is well-converged, with negligible differences among all runs. This is due to the fact that the mass resolution is kept constant across the different simulation boxes.
This convergence suggests that even the smallest simulated volume in the MTNG suite provides a reasonably representative view of the cosmic star formation history.
The overall shape of the SFRD is similar in both the MTNG runs and the observational determinations, featuring a peak in the redshift range $z \approx 2$–$3$ and a general decline toward lower redshift.
The SFRD values in MTNG are also broadly consistent with \citet{2014ARA&A..52..415M} and \citet{2018MNRAS.475.2891D} estimates, though differences by up to a factor of $\lesssim 2$ appear in the redshift range $z \sim 1$–$2$. However, there are important differences worth noting.
First, the position of the SFRD peak differs: it occurs at lower redshift ($z \approx 2$) in the \citet{2014ARA&A..52..415M} determination, whereas the MTNG suite favors a slightly higher redshift value -- the position of the peak is not well-defined in the \citet{2018MNRAS.475.2891D} measurements.
In addition, the cosmic SFRD in \citet{2014ARA&A..52..415M} exhibits a steeper decline toward low redshift relative to both MTNG and \citet{2018MNRAS.475.2891D}. A least-squares fit of the form ${\rm SFRD} \propto (1 + z)^{\kappa_{\rm SFRD}}$ for $z < 1$ yields $\kappa_{\rm SFRD} = 1.74 \pm 0.02$ for MTNG, $\kappa_{\rm SFRD} = 2.26 \pm 0.56$ for \citet{2018MNRAS.475.2891D}, and $\kappa_{\rm SFRD} = 2.58 \pm 0.02$ for \citet{2014ARA&A..52..415M}, with errors reported as 90\% confidence intervals on the slope. 
This quantitative comparison highlights a tension in the low-redshift SFRD evolution between MTNG and the \citet{2014ARA&A..52..415M} results, while showing better agreement with the values from \citet{2018MNRAS.475.2891D}. These differences in redshift evolution may significantly affect the modeling of GW event rates in simulations and impact the interpretation of their evolution based on observational constraints.

    \begin{figure}
        \centering
        \includegraphics[width=0.5\textwidth]{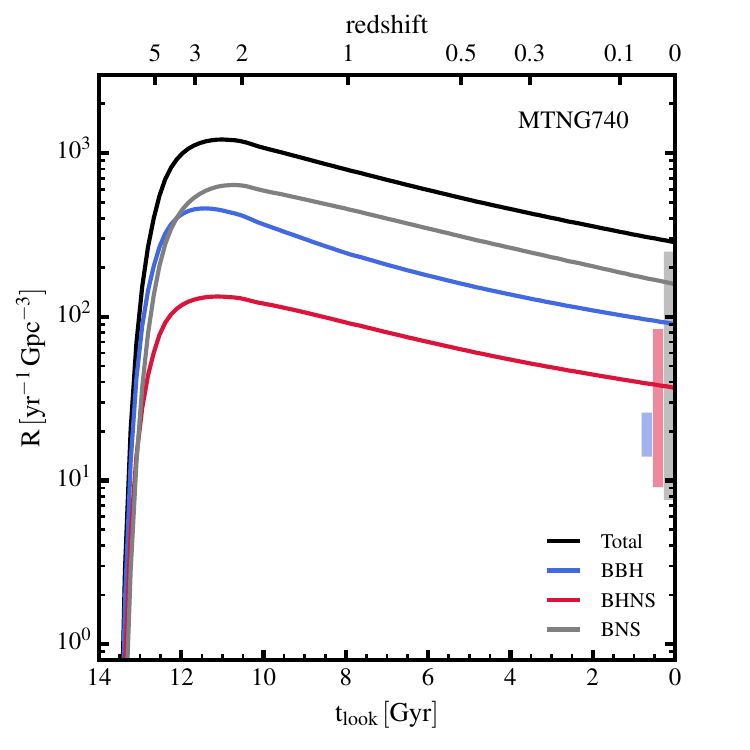}
      \caption{Rate of GW progenitor events per comoving volume as a function of lookback time/redshift in the simulation MTNG740. The total progenitor merger rate and the contribution of different compact binary merger channels are displayed as indicated in the legend. Colored bands represent constraints on the low-redshift ($z\approx0$) merger rate of the different channels, as inferred from GWTC-4 high-confidence events \citepalias{GWTC4}. } 
      \label{fig3}
    \end{figure}

   \begin{figure}
        \centering
        \includegraphics[width=0.5\textwidth]{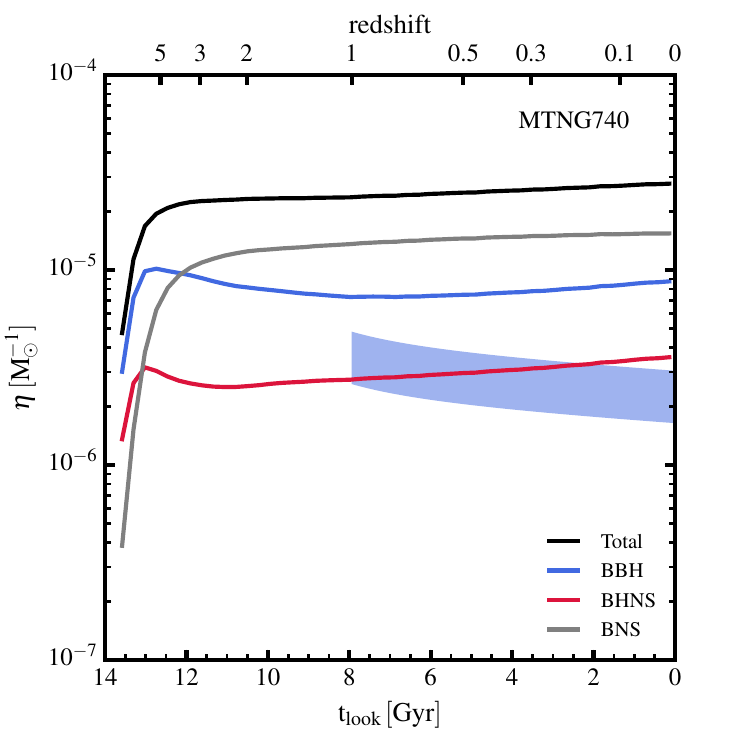}
        
      \caption{Merger efficiency of compact binary objects per unit of stellar mass formed as a function of time/redshift in the simulation MTNG740. The total merger efficiency, along with the contributions from different merger channels, is displayed as indicated in the legend. The shaded region represents constraints on the evolution of BBH merger efficiency at low redshift ($z \lesssim 1$) derived by extrapolating the LVK collaboration limits shown in Fig.~\ref{fig3}. This extrapolation assumes a redshift evolution of the form $(1+z)^{3.2}$ \citepalias{GWTC4}, and adopts the SFRD evolution given by \citet{2014ARA&A..52..415M} reported in Fig.~\ref{fig2}.
    }
         \label{fig4}
   \end{figure}

Figure~\ref{fig3} shows the rate of GW progenitor events per comoving volume as a function of time for the different compact binary merger channels (solid colored lines) and the total rate (solid black line) in the flagship MTNG740 simulation. These quantities are obtained with the same binning procedure used to compute the SFRD presented in Fig.~\ref{fig2}. Also shown are local ($z \approx 0$) rate estimates from the high-confidence events in the GWTC-4 catalog of the LVK collaboration \citepalias[][colored boxes]{GWTC4}.
Thanks to the consistent mass resolution across simulation boxes, the results from other MTNG realizations are essentially unchanged and are presented in Fig.~\ref{fig3b} in \rev{Appendix~\ref{sec:appendixB}}.

We first note that the overall shape of the comoving GW progenitor rates across all three channels is very similar to that of the SFRD. The main difference is in the conversion efficiency of the SFRD into event rates, which is higher for the BNS channel and lower for the BHNS channel. This consistency in shape is expected given the close relationship between the SFRD and the comoving event rate described in eq.~(\ref{eq:rates}).
However, upon closer inspection, some notable differences emerge. Beyond the variation in the efficiency, it is evident that the peak of the BBH and BHNS rates occurs at earlier redshifts ($z \approx 3$). In contrast, the BNS rate peaks at a slightly lower redshift, closer to $z \approx 2$. Both values broadly align with the peak of the SFRD, although the latter lies approximately in the middle of this redshift interval. We speculate that this shift is related to the delay time distributions of the different formation channels. Indeed, the vast majority of progenitor stars in BNS systems typically have very short delay times ($t_{\rm delay} \lesssim 0.1$ Gyr), whereas BBH and BHNS show a non-negligible tail extending towards longer delay times (see Fig.~\ref{fig5} and the related discussion below). 
Another important difference lies in the predicted event rates at low redshift. In particular, the local BHNS and BNS rates from MTNG740 are consistent with current LVK observational constraints -- though these constraints remain relatively weak. In contrast, the BBH channel shows a significant discrepancy: our \sevn-based modeling overestimates the local BBH rate by a factor of $\sim 4.5$. This discrepancy is somewhat reduced -- although still present with the excess reduced to a factor of $\sim 2.4$ -- when the constraints from the previous version of the catalog \citep[GWTC-3;][]{2023PhRvX..13a1048A} are adopted. The overprediction of the local rate of BBH mergers is a known issue with current state-of-the-art binary population synthesis models, particularly when used in combination with a more accurate modeling of the metal-dependent cosmic SFRD, and arises from uncertainties in massive binary star evolution \citep[][]{2024ApJ...976...23B, Sgalletta2025}. In addition, part of the excess in BBH mergers predicted by such models may result from an overestimation of the formation rate of merging BBHs through CE evolution \citep[see, e.g.,][]{Marchant2021}.

\begin{figure}
        \centering
        \includegraphics[width=0.5\textwidth]{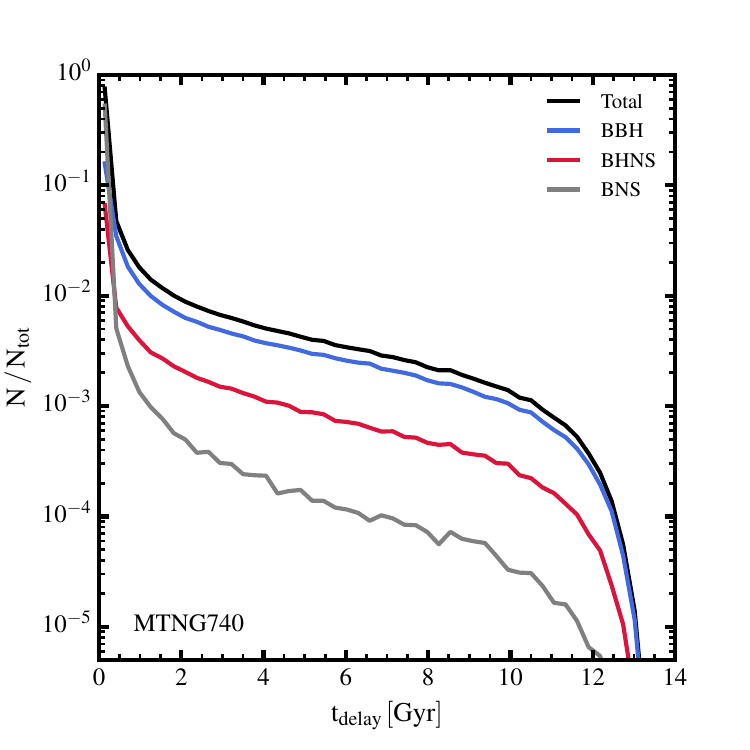}
        
        \caption{Delay time distribution of GW progenitor events in the simulation MTNG740. The delay time distribution of the total merger events, along with the contributions from different compact binary merger channels, is displayed as indicated in the legend. All distributions are normalized to the total number of merger events (i.e., N$_{\rm tot} = $ BBH + BHNS + BNS), so that each delay time bin represents the fractional contribution of that specifc delay time to the overall number of events.
              }
         \label{fig5}
   \end{figure}

Finally, it is worth pointing out that, as with the SFRD, one can characterize the decline toward low redshift of the comoving GW event rate for each channel by measuring its slope as a function of redshift for $z < 1$. Performing a least-squares fit in each case yields $\kappa_{\rm BBH} = 1.44 \pm 0.01$, $\kappa_{\rm BHNS} = 1.34 \pm 0.01$, and $\kappa_{\rm BNS} = 1.55 \pm 0.02$, with errors reported as 90\% confidence intervals on the slope.
In all cases, this decline is shallower than that of the SFRD. This indicates that the merger efficiency, defined as the number of merger events per unit stellar mass formed, tends to increase slightly with time (see also Fig.~\ref{fig4}).
In the case of BBH, we can compare this redshift evolution to observational estimates from \citetalias{GWTC4}, who report a slope of $\kappa_{\rm BBH} = 3.2^{+0.94}_{-1.00}$ for $z \lesssim 1$. Although the  uncertainty on the slope is appreciable, it is evident that the observational data appear to favor a more rapidly declining trend with redshift than the one predicted by our model.

Figure~\ref{fig4} presents the binary merger efficiency per unit stellar mass formed as a function of time for the total GW progenitor events (solid black line) along with individual contributions of the different compact binary merger channels in the MTNG740 box (solid colored lines). The merger efficiency is computed as the ratio between the comoving rate of GW progenitor events (Fig.~\ref{fig3}) and the comoving SFRD (Fig.~\ref{fig2}), thus representing the conversion factor at any given cosmic time $t$ between these two quantities. We also show, as a shaded region, the observational constraints on the BBH merger efficiency at $z \lesssim 1$ obtained by extrapolating the limits displayed in Fig.~\ref{fig3} to higher redshift. This procedure assumes a redshift evolution of the BBH rates $\propto(1+z)^{3.2}$ \citepalias{GWTC4} and adopts the \citet{2014ARA&A..52..415M} SFRD reported in Fig.~\ref{fig2}. Results for all the MTNG boxes analysed in this work are presented in Fig.~\ref{fig4b} in \rev{Appendix~\ref{sec:appendixB}}. We anticipate that also in this case, owing to the consistent mass resolution across boxes, the results from other MTNG realizations are essentially identical.

\begin{figure}
        \centering
       \includegraphics[width=0.5\textwidth]{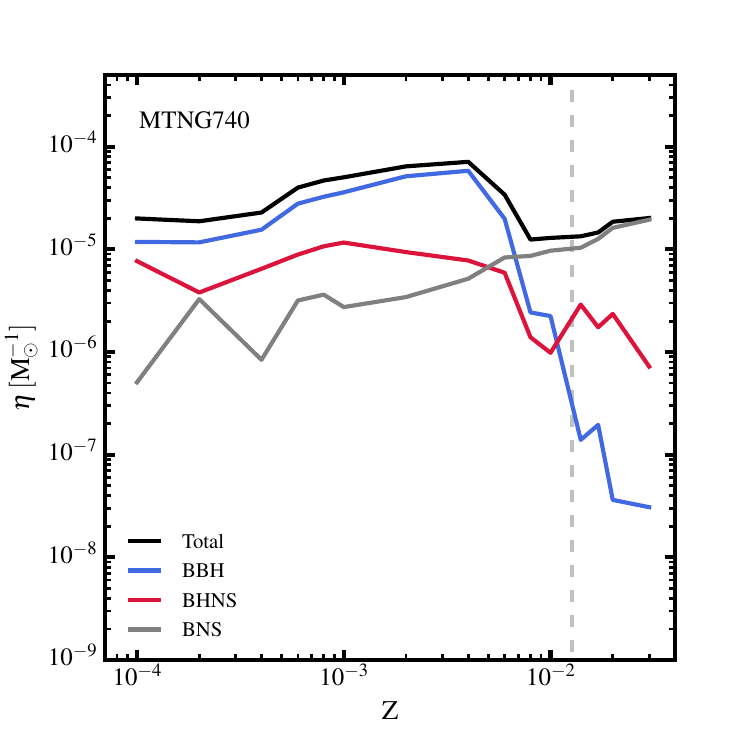}
      \caption{Merger efficiency of compact binary objects per unit of stellar mass formed as a function of the tabulated metallicity of binary systems simulated with the \sevn\ code in the MTNG740 simulation. The total merger efficiency, along with the contributions from different merger channels, is displayed as indicated in the legend. The dashed vertical line indicates the solar metallicity value adopted in MTNG, $Z_\odot = 0.0127$.}
         \label{fig6}
   \end{figure}

All progenitor channels feature a merger efficiency evolution characterized by an initial phase of rapid increase, followed by an approximately constant behavior for $z \lesssim 5$. Below this redshift, variations in efficiency remain within a factor of two for all channels over roughly a Hubble time. This stability further highlights the close connection -- expressed in eq.~(\ref{eq:rates}) -- between star formation and the production of binary mergers responsible for GW signal in the Universe.
Upon closer inspection, some differences in the evolution of merger efficiency among the various channels become apparent. First, the efficiency values differ, reflecting the relative rates of GW progenitor events across channels: BNS mergers exhibit the highest overall rate, followed by BBH and BHNS mergers. A second notable difference lies in the shape of the efficiency evolution. While the BNS efficiency increases monotonically -- albeit slowly -- over time, the BBH and BHNS efficiencies display non-monotonic trend. Specifically, BBH efficiency declines from a local peak at $z \approx 5$ to a minimum at $z \approx 1$, while BHNS efficiency reaches a minimum at $z \approx 3$ before both resume a slow increase toward $z = 0$. These distinct trends are due to a combination of the delay time distribution of merger events (Fig.~\ref{fig5}) and the dependence of event rates on the metallicity of the underlying stellar populations (see Fig.~\ref{fig6} for more details). 

Owing to the current lack of robust observational constraints on the redshift evolution of BHNS and BNS merger rates, it is only possible to compare the efficiency evolution of BBH mergers to observational estimates. The most recent LVK observations \citepalias{GWTC4}, when combined to the SFRD from \citet{2014ARA&A..52..415M}, appear to favor a slowly decreasing efficiency over time for this merger channel. However, it is important to note the considerable uncertainty in the slope of the redshift evolution of the BBH merger rate, $\kappa_{\rm BBH} = 3.2^{+0.94}_{-1.00}$, which implies that the trend predicted by the MTNG results remains somewhat consistent with current observational constraints. Nevertheless, the efficiency values derived from the observations are systematically lower, by a factor of approximately 2.9 to 5.3 at $z = 0$, compared to our results. This tension is sligthly enhanced with respect to earlier LVK constraints obtained from the GWTC-3 catalog \citep[][]{2023PhRvX..13a1048A} at $z=0$, where efficiency values were smaller by factors ranging from 1.2 to 4.7 relative to MTNG predictions.

Figure~\ref{fig5} shows the delay time distribution of GW progenitor events in the MTNG740 simulation, both for the total number of events (solid black line) and for the individual compact binary merger channels (solid colored lines). The distributions are computed using the same binning procedure as in Figs.~\ref{fig2} and \ref{fig3}, and are normalized to the total number of events. As a result, each bin represents the fractional contribution of a specific delay time and merger channel to the overall event population.
The shape of the delay time distribution is broadly similar across all progenitor channels. Most events are concentrated at short delay times ($t_{\rm delay} \lesssim 0.5$ Gyr), with a tail extending to $t_{\rm delay} \sim 12$ Gyr before sharply cutting off. Among the three channels, BBH mergers exhibit the most prominent long-delay tail, followed by BHNS and BNS mergers. This extended tail accounts for only a negligible fraction of BNS events beyond $t_{\rm delay} > 0.5$ Gyr, whereas for BBH and BHNS events, the corresponding fraction rises to $\gtrsim 40\%$.
Moreover, the median delay time for BBH mergers is approximately 0.284 Gyr. In comparison, slightly more than $50\%$ of BHNS events occur within 0.14 Gyr, while for BNS mergers this fraction increases to nearly $96\%$. Similar results are obtained for all the MTNG boxes analysed in this
work and are reported in Fig.~\ref{fig5b} in \rev{Appendix~\ref{sec:appendixB}}. 

These differences in the delay time distribution may help explain the broader spatial extent of BBH mergers relative to BNS events, as shown in Fig.~\ref{fig1}. 
The vast majority of BNS mergers occur within a very short time span ($\lesssim 0.1$ Gyr), making them more tightly linked to recent star formation.
Since star formation is typically concentrated in high-density regions near halo centers, BNS mergers are correspondingly more localized. On the other hand, the longer delay times of BBH mergers allow their progenitors to move farther away from their birth sites, resulting in a more extended spatial distribution. 
The shorter delay times for BNS systems arise from the requirement that they undergo at least one CE phase, which efficiently reduces their orbital separation and leads to faster mergers. In contrast, a fraction of BHNS and BBH systems merge via stable mass transfer on longer timescales \citep[see Fig.~14 in][]{Iorio2023}.

The presence of a more extended tail in the delay time distribution for the BBH and BHNS channels, combined with the declining number of merger events per unit stellar mass formed at metallicities $Z \gtrsim 4 \times 10^{-3}$ (see Fig.~\ref{fig6}), may shed light on the non-monotonic evolution of the binary merger efficiency presented in Fig.~\ref{fig4}. Specifically, the decline in binary merger efficiency for these channels is partially offset by the long delay time events, which contribute to a recovery after the minima observed at $z \approx 3$ (for BHNS) and $z \approx 1$ (for BBH). In contrast, the monotonic increase in the efficiency of the BNS channel can be fully accounted for by the rising number of BNS merger events with increasing metallicity, coupled with the fact that almost all BNS mergers occur within $\sim 0.1$ Gyr after the formation of their parent stellar population.

\subsection{Metallicity dependence of GW progenitor events}

As formalized in eq.~(\ref{eq:rates}), the rates of GW progenitor events can be expressed as a convolution between the SFRD and a channel-dependent factor that encapsulates both the evolving binary merger efficiency per unit stellar mass formed and the delay time distribution of the events. However, eq.~(\ref{eq:rates}) does not explicitly show the dependence of these quantities on metallicity. In fact, it represents a metallicity-integrated version of the underlying components. 
The aim of this section is to  analyze this metallicity dependence and examine how it affects the contributions of different stellar generations to the GW event rates across the BBH, BHNS, and BNS progenitor channels.

In Fig.~\ref{fig6}, we present the merger efficiency of GW progenitor events per unit of stellar mass formed in the MTNG740 simulation as
a function metallicity. The metallicity is binned such that the GW events are assigned to a grid of the binary population synthesis models computed with \sevn, selecting the table that most closely matches the metallicity of their parent star particle.  This is the same table used to stochastically sample the GW event rate for that particle (see Sec.~\ref{sec:GWsampl}). The figure displays both the total efficiency (solid balck line) together with contributions from individual compact binary merger channels (solid colored lines). In all cases, the efficiency is defined as the ratio of the number of GW progenitor events at a given metallicity to the total stellar mass formed at that metallicity. The dashed vertical line is located at $Z = Z_\odot$, with $Z_\odot = 0.0127$ in MTNG  \citep{Asplund2009}. Results from the other MTNG simulations analysed in this work are essentially unchanged with respect the flagship box and are reported in \rev{Appendix~\ref{sec:appendixB}} (Fig.~\ref{fig6b}).

  \begin{figure}
        \centering
        \includegraphics[width=0.5\textwidth]{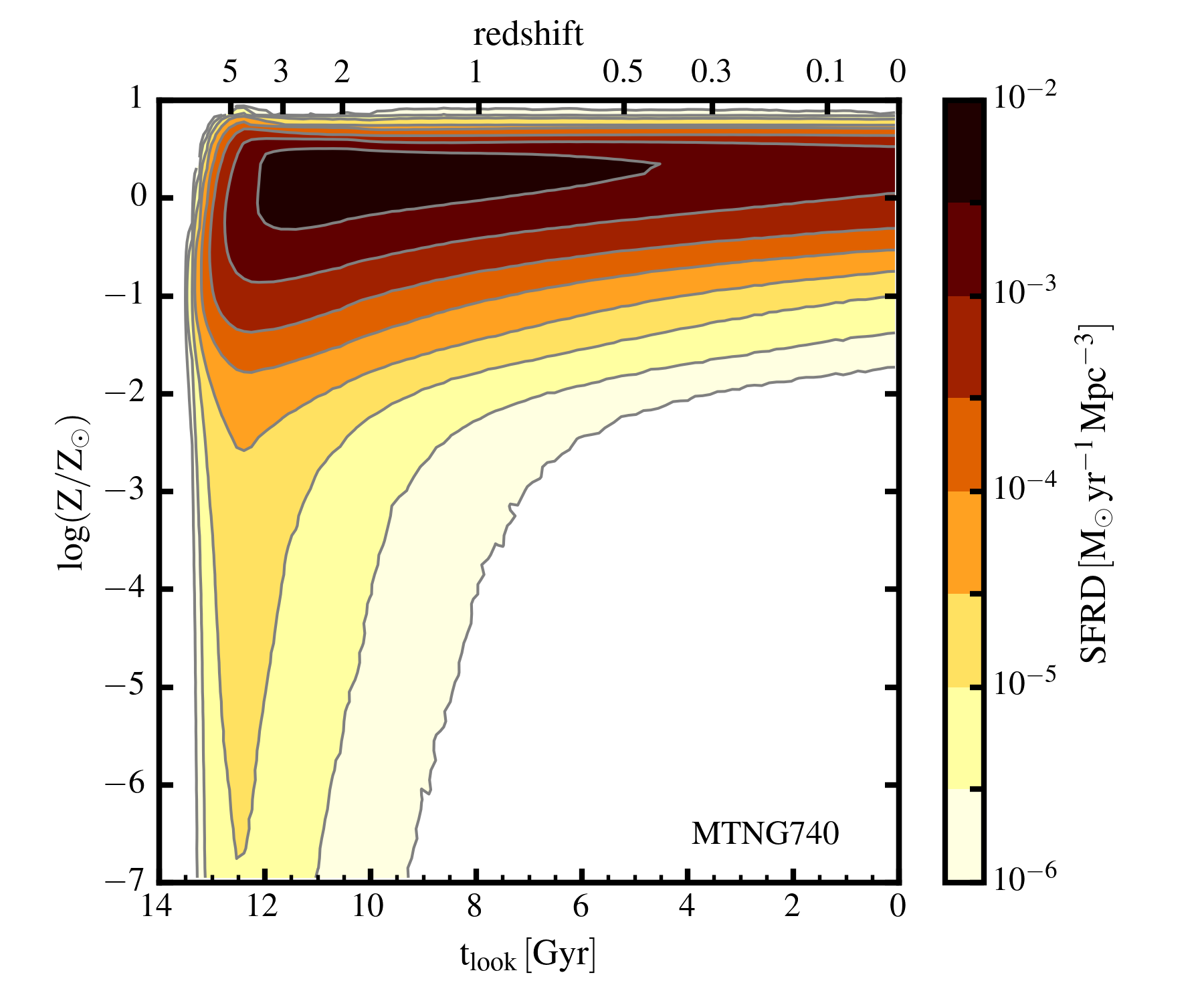}
      \caption{Two-dimensional distribution of metallicity and (lookback) birth times of stellar particles in the MTNG740 simulation. The color scale indicates the corresponding (comoving) star formation rate density. Contours are placed at $[1, 3, 10, 30, 100, 300, 1000, 3000, 10^4] \times 10^{-6} \, {\rm M_\odot \, yr^{-1}\, Mpc^{-3}}$ to help visualize the contributions of metallicity and stellar ages to the overall SFRD of the simulation.
              }
         \label{fig7}
   \end{figure}

The overall trend of the merger efficiency as a function of metallicity begins at $\eta \sim 2 \times 10^{-5}\, \mathrm{M}_{\odot}^{-1}$ at the lowest tabulated metallicity, and steadily increases to a peak of $\eta \sim 6 \times 10^{-5}\, \mathrm{M}_{\odot}^{-1}$ at $Z \approx 4 \times 10^{-3}$. Beyond this point, the efficiency declines, reaching a minimum of $\eta \approx 10^{-5}\, \mathrm{M}_{\odot}^{-1}$ at $Z = 10^{-2}$, followed by a slight rise -- by approximately a factor of two -- at the highest metallicity values considered. This non-monotonic behavior is driven by the varying contributions of different merger channels.
At low metallicities ($Z < 4 \times 10^{-3}$), the BBH channel dominates, contributing roughly an order of magnitude more than the BNS channel and a factor of 3–5 more than the BHNS channel. The decline in the total efficiency at $Z > 4 \times 10^{-3}$ is primarily due to a sharp drop in BBH merger efficiency, from $\eta \sim 4 \times 10^{-5}\, \mathrm{M}_{\odot}^{-1}$ to $\eta \sim 2 \times 10^{-8}\, \mathrm{M}_{\odot}^{-1}$ at $Z = 0.03$. This decrease is partially offset at high metallicities by a steady rise in BNS merger efficiency, which increases by nearly a factor of 30 across the full metallicity range. The BHNS channel exhibits a qualitatively similar trend to BBH, although with generally lower efficiency and a less pronounced decline at high metallicities (approximately a factor of 6 across the range $7 \times 10^{-3} < Z < 10^{-2}$).
Since younger stellar populations tend to be more metal-rich -- because they form from gas that has  been enriched by previous generations of stars -- we expect the BBH and BHNS channels, and particularly the former, to exhibit a pronounced decline in their comoving merger rates at late times. In contrast, due to the increasing merger efficiency of BNS systems with metallicity, the decline in their comoving merger rate at late times is primarily driven by the decreasing SFRD beyond $z \sim 2$. 

However, this simple picture is made more complex by two key points: (i) as discussed in relation to Fig.~\ref{fig5}, GW progenitor events do not occurr instantaneously after the birth of their parent stellar population, but follow a distribution in delay times that can be rather extended -- for instance, in MTNG about $40\%$ of BBH and BHNS mergers occur for $t_{\rm delay} \gtrsim 0.5$ Gyr; (ii) at fixed age, stellar populations can exhibit a relatively broad spread in metallicity, due to local variations in chemical enrichment histories \citep[][]{2005MNRAS.362...41G, 2013A&A...560A.109H, 2020MNRAS.496...80V,
2021MNRAS.503.5826A,2021MNRAS.508.5903V, 
2023MNRAS.523.1565B,
2023ApJ...942...35C, 2025ApJ...981...47G}. These two effects can compensate for the sharp drop of the efficiency at high ($Z > 4 \times 10^{-3}$) metallicity, and soften the declining trend of the comoving merger rates. In the case of BBH mergers, Fig.~\ref{fig3} hints at a flattening of the profile for $z\lesssim 1$, which could potentially result in an overproduction of BBH merger events relative to current LVK constraints. While point (i) arises from the specific modeling choices made in the binary population synthesis simulations used to generate the GW progenitor tables (see Sec.~\ref{sec:method}), point (ii) reflects the chemical enrichment history of stellar populations in the MTNG simulation suite. Chemical enrichment depends on both the star formation history and the adopted stellar yields, which are themselves subject to significant uncertainties. Indeed, discrepancies as large as a factor of two or more are not uncommon among different yield models reported in the literature \citep[see, e.g.,][]{Pillepich2018b, 2023MNRAS.519.3154H}.

To better evaluate the degree of chemical enrichment of the stellar component in the flagship MTNG740 simulation, we show in Fig.~\ref{fig7} the two-dimensional distribution of stellar metallicity versus lookback birth time (i.e., stellar age) for all star particles. This 2D histogram is weighted by the comoving star formation rate density in each bin, computed as the total initial stellar mass formed within the bin, divided by the time bin width and the volume of the simulated box. Results from the other MTNG simulations analyzed in this work are qualitatively similar and are presented in Fig.~\ref{fig7b} in \rev{Appendix~\ref{sec:appendixB}}. The main difference across the boxes is an improved sampling of the low-metallicity, low-age region of the diagram in simulations with larger volumes, due to their higher star particle count.

From Fig.~\ref{fig7}, it is evident that a correlation exists between stellar age and metallicity: in general, younger stars tend to exhibit higher metallicities compared to older ones. In particular, the figure shows that the bulk of star formation in the MTNG740 box occurs at metallicity around $Z \sim Z_\odot$ already by redshift $z \sim 3$, where the cosmic SFRD peaks (see Fig.~\ref{fig2}). Star formation at these metallicities then continues toward $z = 0$, albeit at lower rates. Lower metallicity environments contribute progressively less to the total SFRD over time, with very low metallicities ($Z < 10^{-3}\, Z_\odot$) becoming relevant only at lookback times greater than $\sim$8–9 Gyr. Nevertheless, star formation at relatively low metallicity ($Z \sim 10^{-2}\, Z_\odot$) persists down to the present epoch.

The distribution of the SFRD as a function of metallicity and lookback time helps clarify the evolution of the GW progenitor comoving merger rates shown in Fig.~\ref{fig3}. As illustrated in Fig.~\ref{fig7}, the median metallicity at which most star formation occurs increases rapidly at early times, rising from $\sim 0.1\, Z_\odot$ at $t_{\rm look} \sim 13$ Gyr to approximately solar metallicity at $t_{\rm look} \sim 11$ Gyr. Beyond that point, the median metallicity continues to grow steadily, reaching $Z \sim 1.5\, Z_\odot$ by $z = 0$.
When combined with the metallicity dependence of the binary merger efficiency for each channel (Fig.~\ref{fig6}), this steady rise in the typical metallicity of stellar populations explains the observed timing of the peaks in comoving merger rates. Specifically, the BBH and BHNS rates peak slightly earlier than the SFRD itself (at $z \approx 3$), because their efficiencies decline sharply for $Z \gtrsim 4 \times 10^{-3}$. In contrast, BNS mergers peak later (at $z \approx 2$), because more time is needed for the formation of metal-rich stellar populations, which are more efficient at producing this type of events.

   \begin{figure}
        \centering 
      \includegraphics[width=0.5\textwidth]{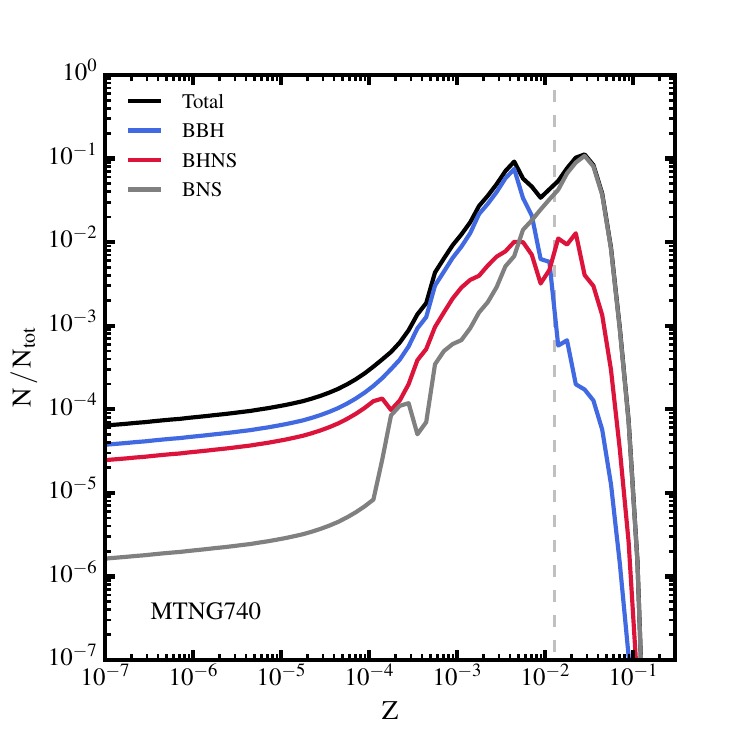}
      \caption{Distribution of GW progenitor events as a function of the metallicity of the parent stellar particle in the MTNG740 simulation. The metallicity distribution of the total merger events, along with the contributions from different compact binary merger channels, is displayed as indicated in the legend. All distributions are normalized to the total number of merger events (i.e., N$_{\rm tot}$ = BBH + BHNS + BNS), so that each metallicity bin represents the fractional contribution of that specific metallicity to the overall number of events. As in Fig.~\ref{fig6}, the dashed vertical line marks the value of $Z_\odot$ adopted in MTNG.
     }
         \label{fig8}
   \end{figure}
   
It is also crucial to quantify the spread in metallicity at fixed stellar age of the SFRD, especially for the BBH and BHNS channels, which are characterized by broader delay time distributions (Fig.~\ref{fig5}). This contrasts with BNS systems, whose mergers typically occur with shorter delays, making their rates more directly coupled to the median metallicity of newly formed stars at a given lookback time. The metallicity spread -- computed as half the width of the 16th–84th percentile range at each lookback time -- is approximately $0.27$–$0.35$ dex in the range $t_{\rm look} \sim 0$–$10$ Gyr, increasing up to $\sim 0.9$ dex beyond $z = 5$. These values indicate that stellar populations with metallicities a factor of $\sim 2$ below the median are common at essentially all cosmic times.
When considering the strong dependence of BBH (and, to a lesser extent, BHNS) binary merger efficiency on metallicity, along with their relatively broad delay time distributions, this spread implies that a non-negligible fraction of low-metallicity progenitors can still contribute to the merger rate even at late times. As previously noted, this delayed component of merger events can reduce the steepness of the decline of BBH merger rates at late times, potentially leading to an overproduction of events relative to current observational constraints.

To conclude our discussion of the metallicity dependence of GW progenitor events, Fig.~\ref{fig8} shows the distribution of the number of progenitors as a function of the metallicity of their parent stellar particle for the MTNG740 simulation. We present both the total distribution of merger events (black solid line) and the contributions from individual compact binary merger channels (solid colored lines). All distributions are normalized to the total number of events in the simulated box and the dashed vertical line marks the location of $Z_\odot$.
Results for the other MTNG boxes analyzed in this work are shown in Fig.~\ref{fig8b} in \rev{Appendix~\ref{sec:appendixB}}. The trends are consistent across all simulated volumes, with the larger volumes providing an improved sampling of the low-metallicity regime due to their higher star particle count.

To some extent, Fig.~\ref{fig8} can be viewed as the ``integrated'' counterpart -- in terms of formed stellar mass at a given metallicity -- of the binary merger efficiency curves presented in Fig.~\ref{fig6}. Consequently, for all channels, we expect a relatively small fraction of merger events at very low metallicities ($Z \lesssim 10^{-4}$), due to the limited star formation occurring in that regime. The varying contributions of different channels at these low metallicities directly reflect the differences in binary merger efficiency (see Fig.~\ref{fig6}).
As metallicity increases above $Z = 10^{-4}$, the total fraction of merger events grows accordingly, following the rise in star formation activity discussed in Fig.~\ref{fig7}. For BBH mergers, the fraction of events peaks at $Z \approx 4 \times 10^{-3}$, followed by a sharp decline driven by the steep drop in merger efficiency at higher metallicities. In contrast, BNS mergers peak at supersolar metallicities ($Z \sim 1.5\, Z_\odot$), which roughly corresponds to the median metallicity where most star formation occurs in the simulation for the last $\sim 10$ Gyr. This is consistent with the BNS merger efficiency, which increases monotonically with metallicity. The subsequent decline in the BNS distribution is primarily due to the small number of stellar particles forming at $Z \gtrsim 1.5 \, Z_\odot$.
BHNS mergers are subdominant compared to BBH at $Z \lesssim 4 \times 10^{-3}$ and to BNS at higher metallicities. Their distribution shows a general behavior similar to the other channels, with a rise in event fraction up to a maximum between $Z \sim 0.004$ and $Z \sim 0.02$, followed by a rapid decline. Interestingly, the BHNS distribution exhibits two distinct peaks -- at $Z \approx 2 \times 10^{-2}$ and $Z \approx 4 \times 10^{-3}$ -- rather than a single maximum. This pattern is consistent with the shape of the merger efficiency for this channel, which shows a more gradual decline at high metallicities compared to BBH, and a mild increase (by a factor of $\sim 2$) in the range $Z = 0.01$–$0.02$.

\subsection{Mass functions of progenitor events and their evolution}

\begin{figure}
        \centering
        \includegraphics[width=0.5\textwidth]{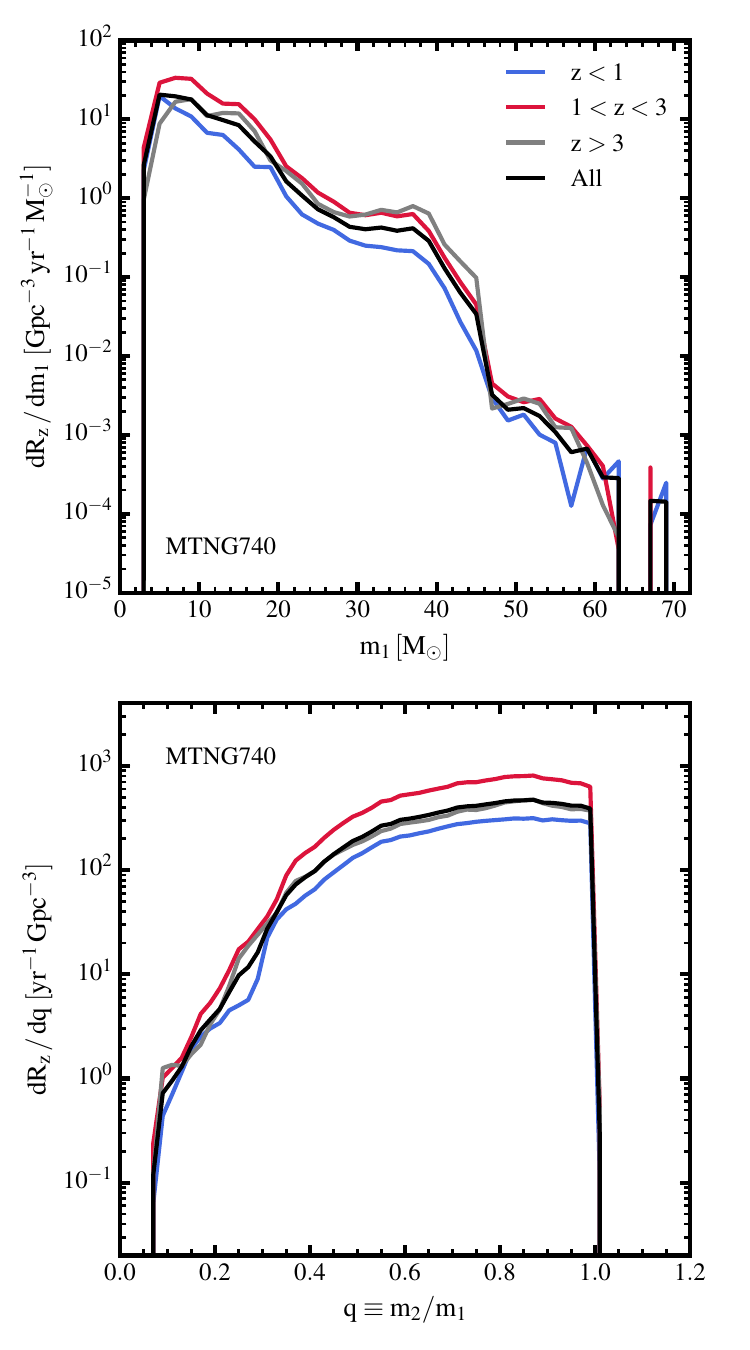}
        
      \caption{Redshift evolution of differential merger rates  of BBH merger GW progenitor events in the MTNG740 box. In the top panel the distributions are displayed as a function of the mass of the primary black hole ($m_1$) of the merging compact binary systems, whereas the bottom \rev{panel} shows the distributions in terms of the mass ratio $q \equiv m_2 / m_1$. Merger rates are normalized relative to the comoving volume of the simulation and to the time span (in Gyr) covered by the redshift intervals reported in the legend.
              }
         \label{fig9}
    \end{figure}

\begin{figure}
        \centering
        \includegraphics[width=0.5\textwidth]{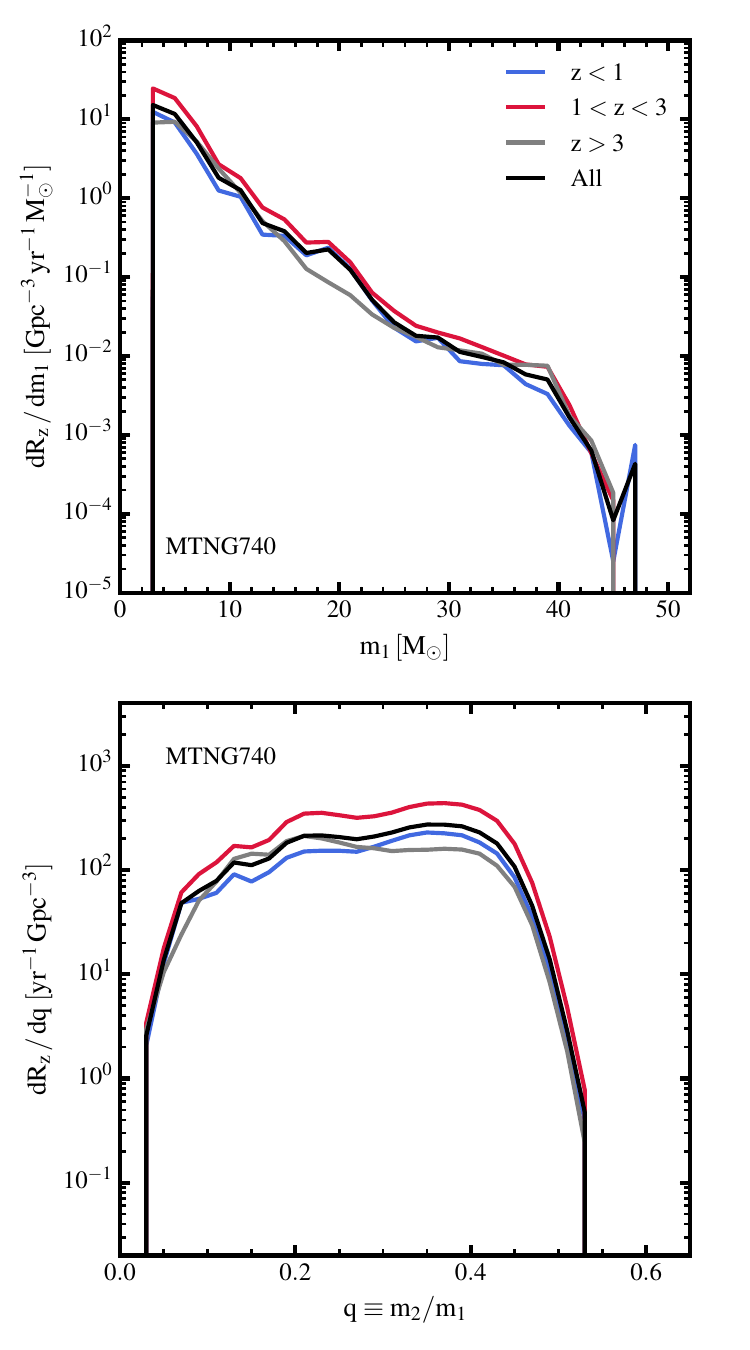}
        
      \caption{The same as Fig.~\ref{fig9}, but for BHNS merger GW progenitor events.
             }
      \label{fig10}
    \end{figure}

Another key property that can be explored using the binary merger catalogues derived from the MTNG simulations is the mass function of progenitor events and its evolution with redshift. We present these mass functions in the MTNG740 box for BBH progenitors in Fig.~\ref{fig9} and for the BHNS progenitors in Fig.~\ref{fig10}. Each figure contains two panels: the top panel shows the mass distribution of the primary object in the compact binary merger ($m_1$), while the bottom panel displays the distribution of the mass ratio ($q \equiv m_2 / m_1$) of the binary system.
Mass functions are shown for the entire redshift range covered by the simulation (black lines), as well as for three distinct redshift intervals: $z < 1$ (blue), $1 < z < 3$ (red), and $z > 3$ (gray). All distributions are normalized to the comoving volume of the simulation box and to the duration (in Gyr) of the respective redshift interval. Since the mass of BNS systems is resampled from a Gaussian distribution to reproduce the Galactic binary neutron star population (see Sec.~\ref{sec:SEVN}), we do not discuss its properties here.
Results from the other MTNG simulation boxes are consistent with those of the flagship MTNG740 run and are shown in \rev{Appendix~\ref{sec:appendixB}} in Fig.~\ref{fig9b} for BBH progenitors and Fig.~\ref{fig10b} for BHNS progenitors. For completeness, \rev{Appendix~\ref{sec:appendixB}} also presents the mass functions of the BNS merger channel in Fig.~\ref{fig11b}. 

For BBH mergers (Fig.~\ref{fig9}), it is evident that the mass distribution -- both in terms of the primary mass and the mass ratio -- evolves only mildly with redshift. The main variation lies in the normalization of the distributions, while their overall shape remains largely unchanged. These differences in normalization reflect the evolution of the average comoving BBH merger rates across cosmic time (see Fig.~\ref{fig3}).
Specifically, the lowest normalization is observed at $z < 1$, consistent with the monotonic decline in the BBH comoving merger rate at late times. In contrast, higher redshifts, particularly the $1 < z < 3$ interval where the merger rate peaks (around $z \approx 3$), show the largest values of the mass distribution. The distribution computed over the entire simulation timespan falls between these two extremes, as expected. The mild evolution of the BBH mass function with redshift is consistent with the findings of \citet{2019MNRAS.487....2M}, who performed a similar analysis -- albeit with more numerous and less wide redshift bins -- using the Illustris-1 simulation. Notably, their study employed a comparable methodology to generate GW progenitor catalogues, though based on an earlier version of the binary population synthesis model (\textsc{mobse}) and applied to a cosmological simulation with different galaxy formation physics prescriptions.

The BBH mass distribution exhibits a steadily decreasing trend with the mass of the primary black hole, spanning from $m_1 \approx 4\, \mathrm{M}_\odot$ up to  $m_1 \approx 70\, \mathrm{M}_\odot$. A particularly notable feature is a sharp drop at $m_1 \simeq 44\, \mathrm{M}_\odot$, where the mass functions  decreases by about two orders of magnitude. This is a consequence of the complex interplay between pair-instability supernovae and envelope stripping in binary star evolution, which produce a plateau at around $30-40$ $\mathrm{M}_\odot$, and binary evolution processes -- such as CE phase and mass transfer -- that remove mass from the system, hindering the formation of BHs with masses $\gtrsim 50$ $\mathrm{M}_\odot$ \citep{2021MNRAS.505.3873B,Iorio2023}. In agreement with the latest observational constraints from LVK \citepalias[][their Fig.~3; see also Fig.~11 of \citealt{2023PhRvX..13a1048A} for results obtained with the previous version of the GWTC catalog]{GWTC4}, the distribution shows local maxima around $m_1 \approx 10\, \mathrm{M}_\odot$ and $m_1 \approx 36\, \mathrm{M}_\odot$, although the prominence of the latter varies across redshift bins. Our results also indicate that the distribution extends to lower primary masses compared to LVK estimates. The absence of BBH merger progenitors with $m_1 \gtrsim 70 \,\mathrm{M}_\odot$ is likely a consequence of not including the dynamical merger channel, which can produce systems with such high primary masses \citep[e.g.,][]{Mapelli2016,Rodriguez2016,DiCarlo2020}, in our models . Regarding the mass ratio distribution, we observe an opposite trend: the differential merger rate increases with $q$,  rising by approximately two orders of magnitude in the range $0.2<q<0.4$ and then continuing to grow more gradually up to $q \sim 1$. This indicates a preference for BBH mergers involving near-equal mass components and may be a signature of mergers predominantly driven by a CE envelope evolutionary phase. However, near-equal mass BBH mergers in \sevn\ typically arise from a combination of binary evolution processes and supernova models \citep[see][their Fig.~14]{Iorio2023}.

A mild evolution of the mass distribution with redshift is also found for the BHNS channel (Fig.~\ref{fig10}). In this case, the mass distribution of the primary component -- i.e., the black hole -- exhibits a shape broadly similar to that of BBH mergers. The distribution decreases monotonically with increasing $m_1$, spanning a narrower mass range from $m_1 \approx 4\, \mathrm{M}_\odot$ to $m_1 \approx 48\,\mathrm{M}_\odot$. A noticeable feature is a bump around $m_1 \approx 20\,\mathrm{M}_\odot$ for mergers occurring at redshifts $z < 3$.
The distribution of BHNS mergers as a function of the mass ratio $q$ is approximately flat for $q \gtrsim 0.2$ up to $q \simeq 0.45$, beyond which it drops sharply. This decline at high $q$ values is expected, given that the neutron star is significantly less massive than its black hole companion. An exception to this trend is observed in the $z > 3$ redshift interval, where the flat portion of the distribution begins at $q \gtrsim 0.1$. The normalization trends across different redshift ranges are consistent with those found for the BBH  channel.
The relative flatness of the $q$ distribution suggests that BHNS mergers occur over a wide range of mass ratios, without a strong preference for any particular value of $q$. This reflects the broader diversity in mass configurations characteristic of these mixed compact object binaries.

\section{Summary and conclusions}\label{sec:conclusion}

In this paper, we introduced a flexible and fully parallel framework to build GW event catalogs in full-physics simulations of galaxy formation. Our framework works by associating binary merger events across three different channels (BBH, BHNS,  and BNS) to stellar particles representing stellar populations in cosmological  simulations. This association is performed by \arepogw\ -- a module integrated in the moving-mesh code \arepo\ -- by stochastically sampling (based on the mass, age and metallicity of the underlying stellar population) binary mergers tables obtained with the state-of-the-art population-synthesis code \sevn. \arepogw\ can operate in two main modes: (i) on-the-fly, performing the association of merger events during the course of a (hydrodynamical) galaxy formation simulation, and (ii) in post-processing, using as input a snapshot from an existing run. Our framework can be easily adapted to work with simulation outputs written in formats other than the standard \arepo\ format, as it requires only minimal input -- namely, the initial mass, metallicity, and age of the star particles.
Moreover, it also offers broad applicability, as it can be used with different sets of binary merger tables produced by \sevn\ or other stellar population synthesis codes.

To demonstrate the capabilities of  our new framework, we apply \arepogw\ in its post-processing mode to the MTNG simulation suite, one of the largest and most advanced cosmological simulation projects to date.
We build GW progenitor events catalogs for all four simulation boxes in the MTNG suite, including the flagship MTNG740 run, yielding a total data volume of $\approx30$ TB.
In this work, we focused on the key global properties of the GW events (for instance, their comoving rates, formation efficiencies, delay time distributions, and the metallicity and mass distributions of their progenitors) and examined how these properties vary as a function of the simulated volume. Our main results can be summarized as follows:

\begin{enumerate}[(i)]
\item the comoving rate of GW progenitor events follows the evolution of the cosmic SFRD, presenting a peak at early cosmic times ($z\approx 3$ for BBH and BHNS and $z \approx 2$ for BNS) followed by a decline toward $z=0$. BBH and BHNS events tend to peak slightly earlier than the SFRD, whereas BNS events more closely track it. The decline in the comoving rate for all GW progenitor channels is somewhat shallower than that of the SFRD, suggesting an increasing merger efficiency per unit stellar mass formed at lower redshifts;

\item the rates of BHNS and BNS at $z\sim 0$ are consistent with recent LVK observational determinations. In contrast, BBH rates are overproduced by a factor of $\sim 4.5$, a discrepancy also seen in other predictions based on different stellar population synthesis models. In particular, LVK observations seem to favor a steeper slope of the redshift evolution of the BBH merger rate for $z \lesssim 1$ compared to what is found in MTNG, although there is still a considerable uncertainty in the slope derived from observations;

\item GW progenitor event rates are shaped by two key factors: (i) their delay time distribution, which strongly depends on the assumptions made in stellar population synthesis modeling;
(ii) the metallicity dependence of merger efficiency, which varies across channels and reflects the chemical enrichment history of the simulation. 
In MTNG, the quite rapid metal enrichment of stellar populations, together with a declining merger efficiency at higher metallicity for BBH and BHNS systems, explains their earlier peak relative to the SFRD. The narrow delay time distribution of BNS events ($t_{\rm delay} \lesssim 0.1$ Gyr) results in a redshift evolution closely linked to recent star formation. Conversely, the long-delay tail of BBH events may contribute to their excess at late times relative to current constraints from the LVK collaboration;

\item we also examined the mass functions of the different GW progenitor channels, both in terms of the distribution of primary binary masses ($m_1$) and merger mass ratios ($q$), as well as their evolution with redshift. In the three redshift bins that we analysed we found only a mild evolution of the mass functions, in agreement with previous studies that adopt different binary population synthesis models and galaxy formation prescriptions. In particular, the shape of the mass functions remains largely unchanged with redshift, with only modest differences in normalization, consistent with the expected redshift evolution of the merger rates. This mild evolution is driven by the rapid metal enrichment of stellar populations in MTNG, which reach metallicities of $Z \sim Z_\odot$ by $z\sim 3$ and then  stay roughly at the same levels down to $z = 0$. This in turn implies that the mass distributions of binary mergers remain nearly unchanged over time;
 
\item the results discussed above remain consistent across all four MTNG simulation boxes analyzed in this work. This is not surprising, as the simulations cover progressively larger volumes of the Universe while maintaining the same mass resolution. Consequently, converged results for the SFRD are obtained across different boxes, leading to similar levels of metal enrichment and, therefore, comparable statistics for GW progenitor properties. It is worth noting that the MTNG model is based on the IllustrisTNG galaxy formation framework, which has demonstrated good convergence properties in general. However, drifts of up to $\sim 30-40\%$ in stellar mass with resolution have been observed, particularly for low-mass galaxies \citep{Pillepich2018}. Therefore,  SFRs for low-mass galaxies are numerically biased low in MTNG, so that some impact on the amplitude of the event rates can be expected at higher resolution. 
\end{enumerate}

The analysis presented in this paper represents a first step in exploiting the GW event catalogues produced for the MTNG simulations with \arepogw\ in conjunction with \sevn\ predictions. In particular, full light-cone outputs are available for MTNG, which can be combined with the GW event catalogues to generate light-cone outputs for GW events. This type of output will enable studies of their spatial distribution, clustering properties, correlations, and bias relative to the underlying stellar and matter density fields across multiple angular scales.

Another promising avenue made possible by the combination of the MTNG cosmological simulations and our GW event catalogue is to associate the properties of the GW signals with those of the host galaxies of the events progenitors. This would allow investigations into how GW signals are influenced by local galaxy properties and the environments in which these objects reside. Such insights could also facilitate the implementation of a method similar to the one presented in this work within semi-analytical models of galaxy formation \citep[e.g.,][]{2015MNRAS.453.4337S,2016MNRAS.461.1760H,2016MNRAS.462.3854L,2020MNRAS.491.5795H}.
Semi-analytical models offer a computationally cheaper approach than full-physics simulations for exploring how the initial properties of simulated binary systems and the modeling of key physical processes driving binary evolution -- such as those adopted in \sevn\ or other stellar population synthesis codes -- affect compact binary merger rates. The accuracy of these underlying assumptions represents a significant source of uncertainty in the results presented here. 

\rev{Indeed, a current limitation of \arepogw\ is that only one binary population synthesis model can be used at a time. However, this does not diminish the value of applying the framework to large-scale cosmological simulations. Variations in galaxy formation physics can lead to substantial differences in stellar masses, star formation histories, and metallicity distributions, all of which directly affect the predicted population of GW progenitor events even when the same binary evolution model is used. Because different galaxy formation models must be run as separate simulations in any case, the on-the-fly mode offers an effective means to capture these differences self-consistently. By contrast, exploring the parameter space of binary evolution models remains computationally more efficient in post-processing, since a single hydrodynamical simulation can be reused. Future versions of \arepogw\ may support multiple population synthesis models or multiple independent samplings of the same model. Such extensions are expected to require only a moderate additional memory cost and would enable more comprehensive uncertainty quantification while keeping the framework computationally feasible.}

Finally, starting from our GW event catalogue, it would be extremely informative to produce mock observations of GW {signals} that fully incorporate observational uncertainties and selection effects, in order to enable more robust comparisons with, and help the interpretation of, actual GW data. We will explore these research directions in future work, thereby advancing our understanding of gravitational wave {signals} and their connection to galaxy formation.

\begin{acknowledgements}
We thank the MTNG collaboration for kindly providing advance access to the simulation data.
FM and MB gratefully acknowledge funding from the European Union -- NextGenerationEU, in the framework of the HPC project – “National Centre for HPC, Big Data and Quantum Computing” (PNRR -- M4C2 -- I1.4 -- CN00000013 -– CUP J33C22001170001). GI was supported by a fellowship grant from the la Caixa Foundation (ID 100010434) under grant agreement LCF/BQ/PI24/12040020. MCA acknowledges financial support from  FONDECYT Iniciaci\'on 11240540 and ANID BASAL project FB210003. MM acknowledges financial support from the Deutsche Forschungsgemeinschaft (DFG, German Research Foundation) under Germany’s Excellence Strategy EXC 2181/1 -- 390900948 (the Heidelberg STRUCTURES Excellence Cluster) and from the European Research Council for the ERC Consolidator grant DEMOBLACK, under contract no. 770017. SB is supported by the UK Research and Innovation (UKRI) Future Leaders Fellowship [grant number MR/V023381/1]. VS and LH acknowledge support from the Simons Foundation through
the ``Learning the Universe'' initiative.
The initial conditions for the binary population synthesis simulations were generated with  the {\sc IC4popsyn} code (\href{https://github.com/GiacobboNicola/IC4popsyn/tree/main}{https://github.com/GiacobboNicola/IC4popsyn/tree/main}).
\sevn\ simulations used in this work were made possible through a CINECA-INFN (TEONGRAV) agreement, providing access to resources on Leonardo at CINECA. 
For the analysis and visualization of simulation data we used the Python packages NumPy \citep{Harris2020}, SciPy \citep{Virtanen2020}, Matplotlib \citep{Matplotlib}, and mpi4py \citep{Dalcin2005}. The analysis presented in this paper was performed on the HPC computing cluster of the Open Physics Hub project (\href{https://site.unibo.it/openphysicshub/en}{https://site.unibo.it/openphysicshub/en}) hosted by the Department of Physics and Astronomy at the University of Bologna.
\end{acknowledgements}

\bibliographystyle{aa}
\bibliography{biblio_gw}

\onecolumn

\begin{appendix}
\section{Quantitative comparison between the on-the-fly and post-processing modes of \arepogw}\label{sec:appendixA}
\rev{As discussed in Sec.~\ref{sec:GWsampl}, the two operational modes of \arepogw\ process the stellar particles of a specific simulation using a slightly different approach. Given that the association between star particles and GW progenitor events is performed stochastically in both modes, it is not possible to obtain the exact same sequence of binary merger events even when analysing the same simulation. The goal of this Appendix is to compare the on-the-fly and post-processing modes and quantify the extent of the resulting discrepancies.}

\rev{For this comparison, we decided to rerun the MTNG46 box with the on-the-fly mode of \arepogw\ enabled and postprocess the redshift zero snapshot of this new simulation for consistency. This setup enables a direct, controlled comparison between the two GW catalogues produced with the two approaches, ensuring that the same set of stellar particles (with identical formation masses, formation times, and metallicities) is used in both cases. Therefore, any differences in the outcome can be attributed solely to the operational mode of \arepogw.}

\begin{figure}[h!]
        \centering
        \includegraphics[width=0.95\textwidth]{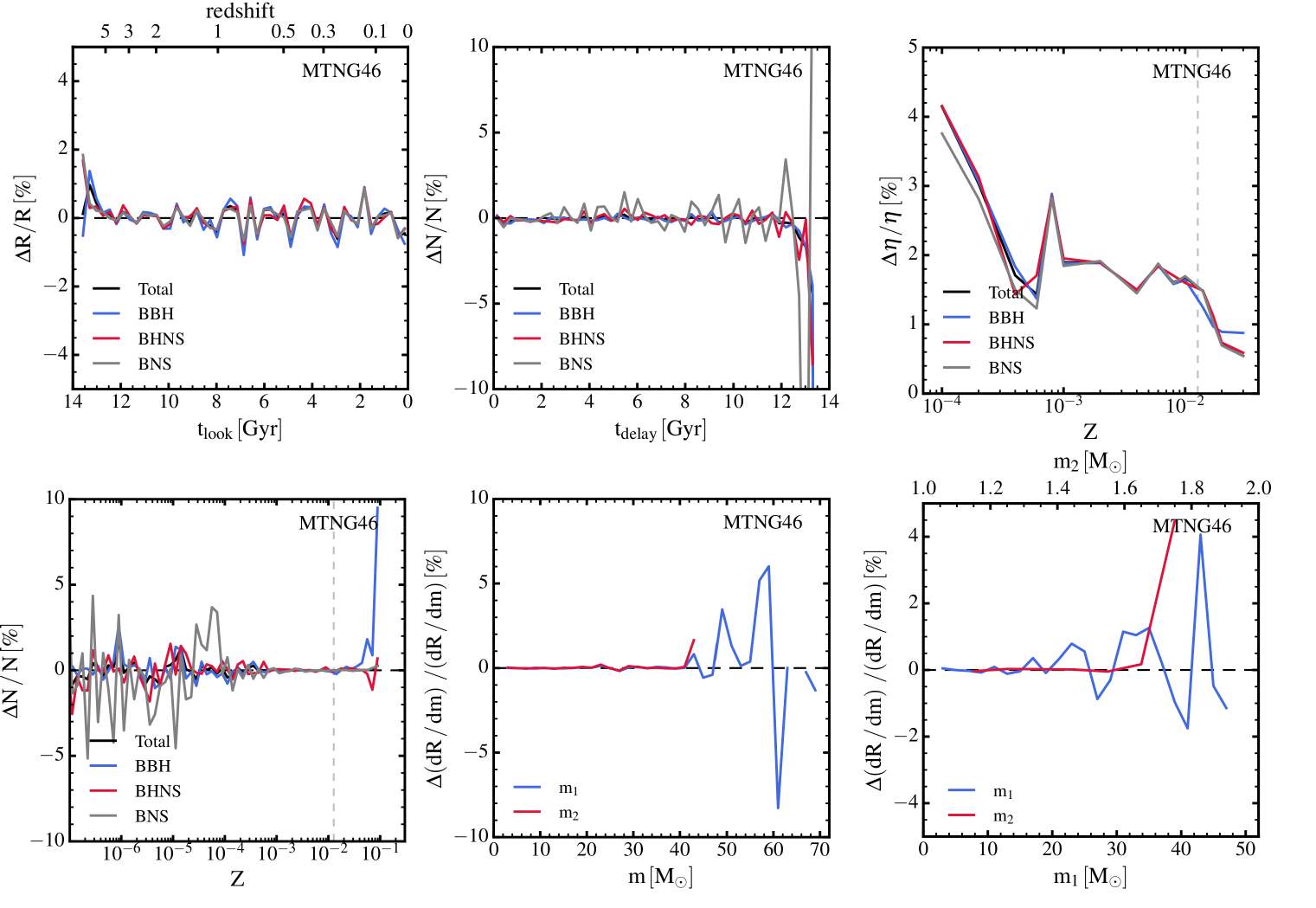}
      \caption{\rev{Relative differences between properties of GW progenitor events obtained with the on-the-fly and post-processing modes of \arepogw\ for a rerun of the MTNG46 simulation. Differences are expressed as percentages, with the post-processing mode taken as the reference case. Top row: Comoving rates as a function of time (left panel), delay time distribution (middle panel), and merger efficiency as a function of tabulated \sevn\ metallicity (right panel). Bottom row: distribution of GW progenitor events as a function of the metallicity of the parent star particle (left panel), BBH (middle panel) and BHNS (right panel) mass functions}.}
         \label{fig3a}
\end{figure}

\rev{Below, we report the results of this analysis. We present a comparison of the comoving rate of GW progenitor events as a function of lookback time (Fig.~\ref{fig3a}, top left panel), the delay time distribution of GW progenitor events (Fig~\ref{fig3a}, top middle panel), the merger efficiency of compact binary objects as a function of the tabulated metallicity of binary
systems (Fig~\ref{fig3a}, top right panel), the distribution of GW progenitor events as a function of the metallicity of the parent stellar particle (Fig~\ref{fig3a}, bottom left panel), and the BBH and BHNS mass functions (Fig~\ref{fig3a}, bottom middle and bottom right panel, respectively). For each quantity, the panels show the relative difference between the two approaches, adopting the post-processing mode as the reference case. These results clearly demonstrate that the impact of the specific operational mode of \arepogw\ on the statistical properties of the population of the GW progenitor events is minimal. The observed differences are at most of the order of a few percent  and can be attributed to the stochastic nature of the association procedure between simulated star particles and compact binary mergers implemented in \arepogw. Occasionally, larger relative differences -- up to $\sim 10-20\%$ -- can be present, but only for rare events, such as systems with long delay times ($\gtrsim 12$ Gyr) or extremely metal-rich ($Z \gtrsim 0.1$) or high-mass ($m_1 \gtrsim 55\,{\rm M_\odot}$) progenitors. Larger simulation boxes are expected to yield comparable or more consistent results between the two modes, due to the increased number (and therefore better sampling) of GW progenitor events.}

\section{GW progenitor properties across different MTNG boxes}\label{sec:appendixB}

In this Appendix, we present the same set of results shown in the main body of the paper for the flagship MTNG740 simulation, extended to all the boxes of the MTNG suite for which we produced a GW progenitor catalogue. As mentioned above, the four analyzed boxes differ only in terms of the simulated volume, but adopt the same prescriptions and parameterization for galaxy formation physics, as well as the same mass resolution. 

\begin{figure*}[h!]
        \centering
        \includegraphics[width=\textwidth]{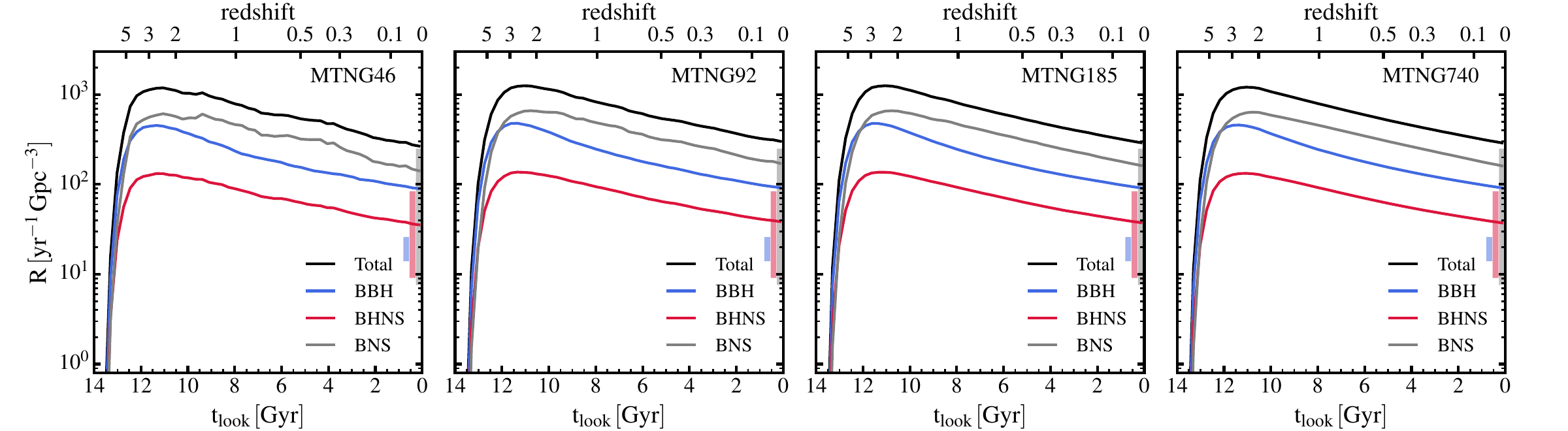}
      \caption{Rate of GW progenitor events per comoving volume as a function of lookback time/redshift in four different boxes of the MTNG simulation suite: MTNG46, MTNG92, MTNG185, and MTNG740 (from left to right). The total progenitor merger rate and the contribution of different compact binary merger channels are displayed as indicated in the legend. Colored bands represent constraints on the low-redshift ($z \approx 0$) merger rate of the different channels, as inferred from GWTC-4 high-confidence events \citepalias{GWTC4}.
    }
         \label{fig3b}
   \end{figure*}

\begin{figure*}[h!]
        \centering
        \includegraphics[width=\textwidth]{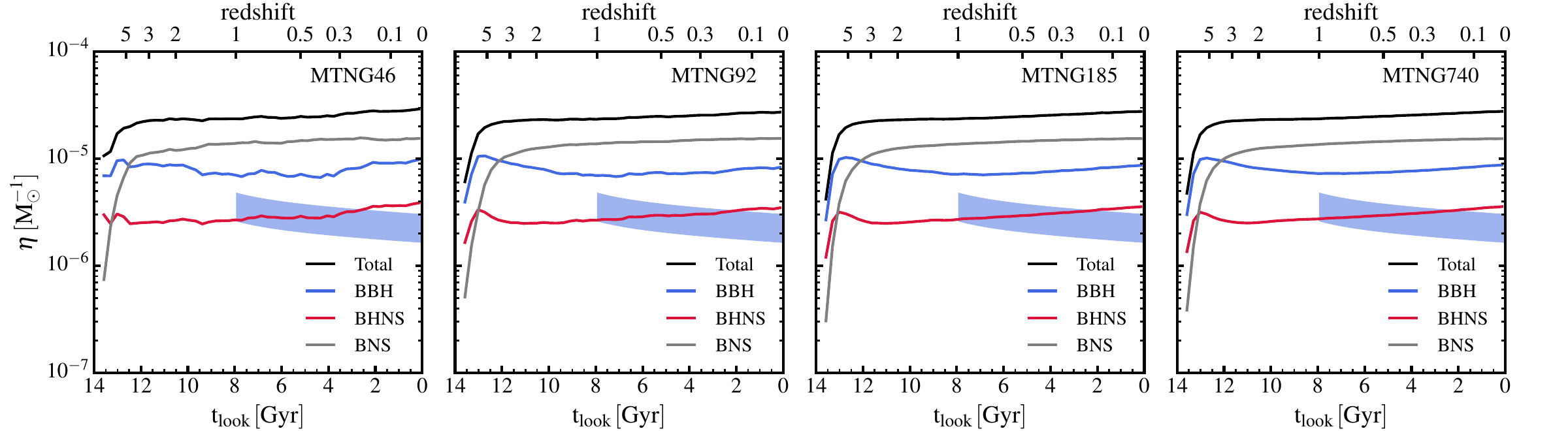}
      \caption{Merger efficiency of compact binary objects per unit of stellar mass formed as a function of time/redshift in four different boxes of the MTNG simulations: MTNG46, MTNG92, MTNG185, and MTNG740 (from left to right). The total merger efficiency, along with the contributions from different merger channels, is displayed as indicated in the legend. Shaded regions represent constraints on the  evolution of BBH merger efficiency at low redshift ($z \lesssim 1$), derived by extrapolating the \citetalias{GWTC4}  limits as in Fig.~\ref{fig4}.
              }
         \label{fig4b}
   \end{figure*}

\begin{figure*}
        \centering
        \includegraphics[width=\textwidth]{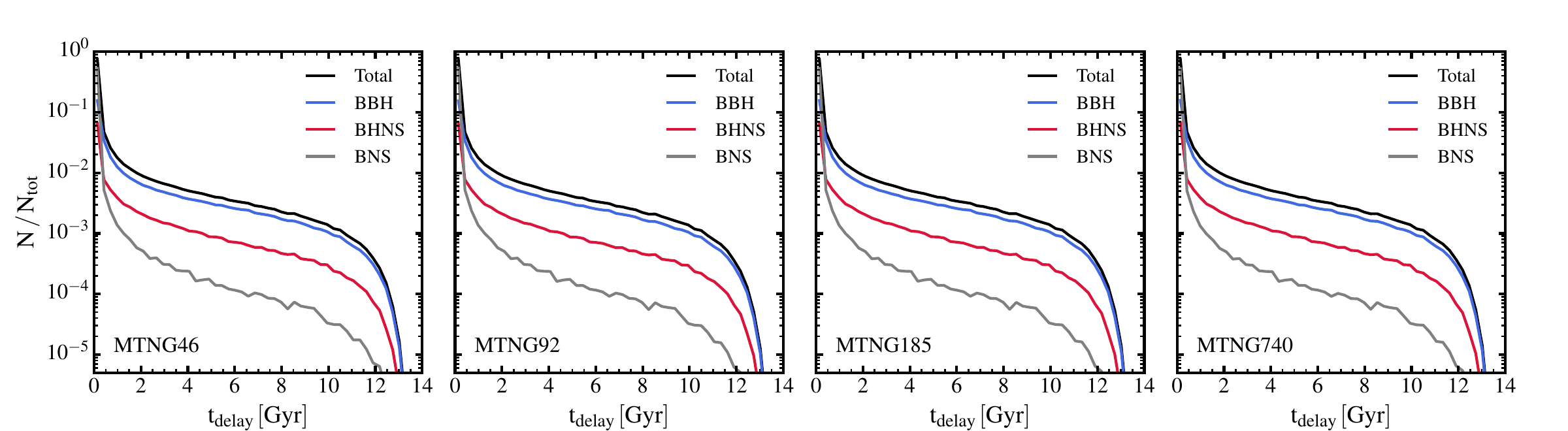}\caption{Delay time distribution of GW progenitor events in four different boxes of the MTNG simulations: MTNG46, MTNG92, MTNG185, and MTNG740 (from left to  right). The delay time distribution of the total merger events, along with the contributions from different compact binary merger channels, is displayed as indicated in the legend. All distributions are normalized to the total number of merger events.
              }
         \label{fig5b}
   \end{figure*}

\begin{figure*}
        \centering
        \includegraphics[width=\textwidth]{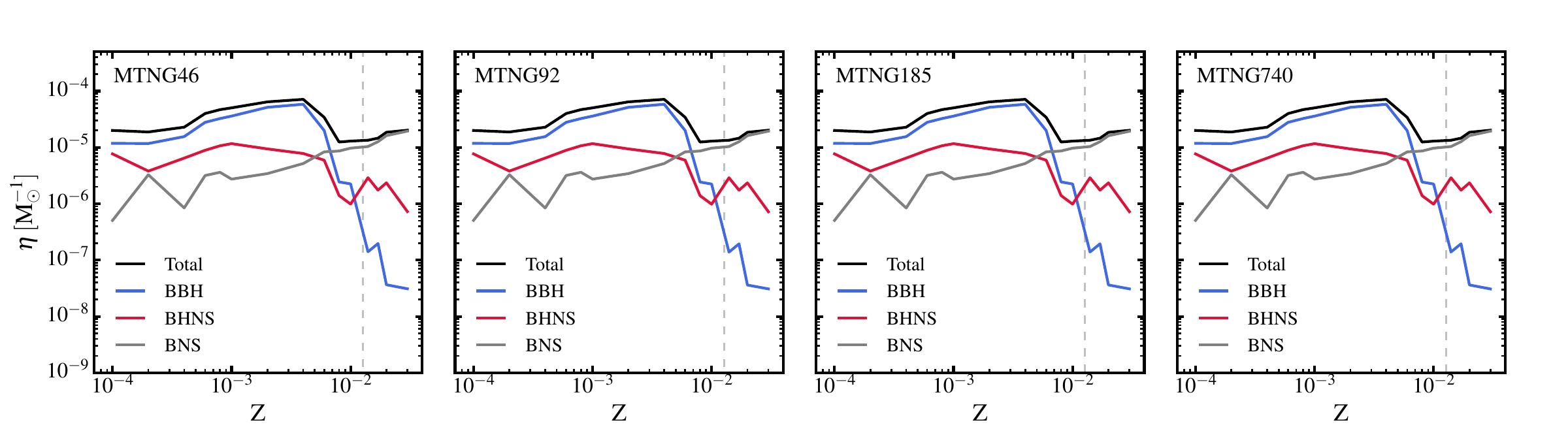}
      \caption{Merger efficiency of compact binary objects per unit of stellar mass formed as a function of tabulated metallicity of binary systems simulated with the \sevn\ code in four different boxes of the MTNG simulations: MTNG46, MTNG92, MTNG185, and MTNG740 (from left to right). The total merger efficiency, along with the contributions from different merger channels, is displayed as indicated in the legend. The dashed vertical lines indicate the solar metallicity value, $Z_\odot = 0.0127$, adopted in MTNG.
              }
         \label{fig6b}
   \end{figure*}

\begin{figure*}
        \centering
        \includegraphics[width=\textwidth]{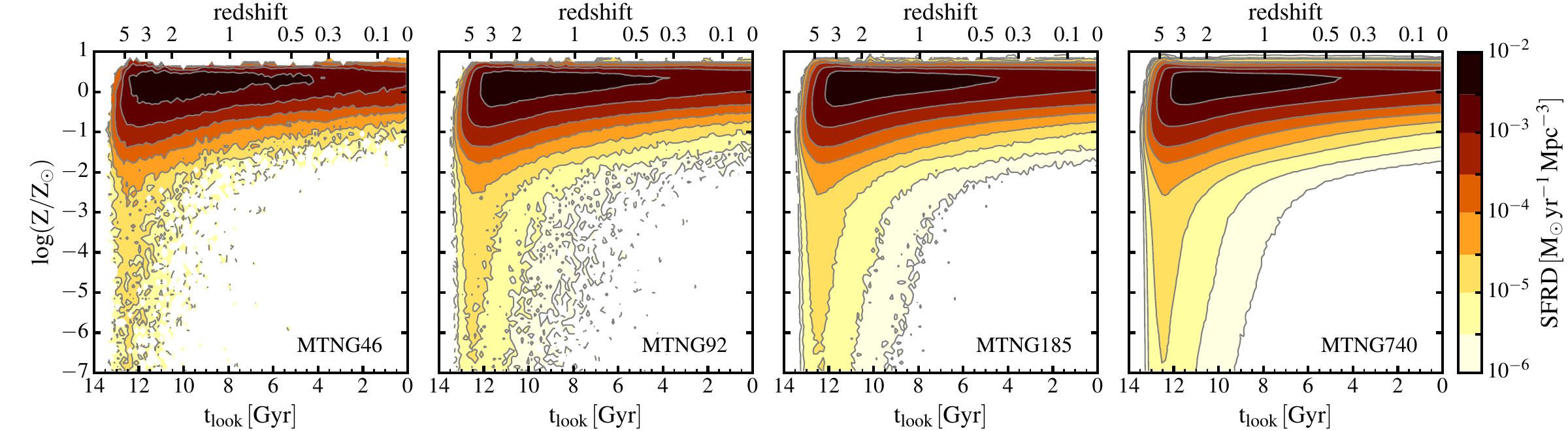}
      \caption{Two-dimensional distribution of metallicity and (lookback) birth times of stellar particles in four different boxes of the MTNG simulations: MTNG46, MTNG92, MTNG185, and MTNG740 (from left to right). The color scale indicates the corresponding (comoving) star formation rate density. Contours are placed at $[1, 3, 10, 30, 100, 300, 1000, 3000, 10^4] \times 10^{-6} \, {\rm M_\odot \, yr^{-1}\, Mpc^{-3}}$.
              }
         \label{fig7b}
   \end{figure*}

\begin{figure*}
        \centering 
        \includegraphics[width=\textwidth]{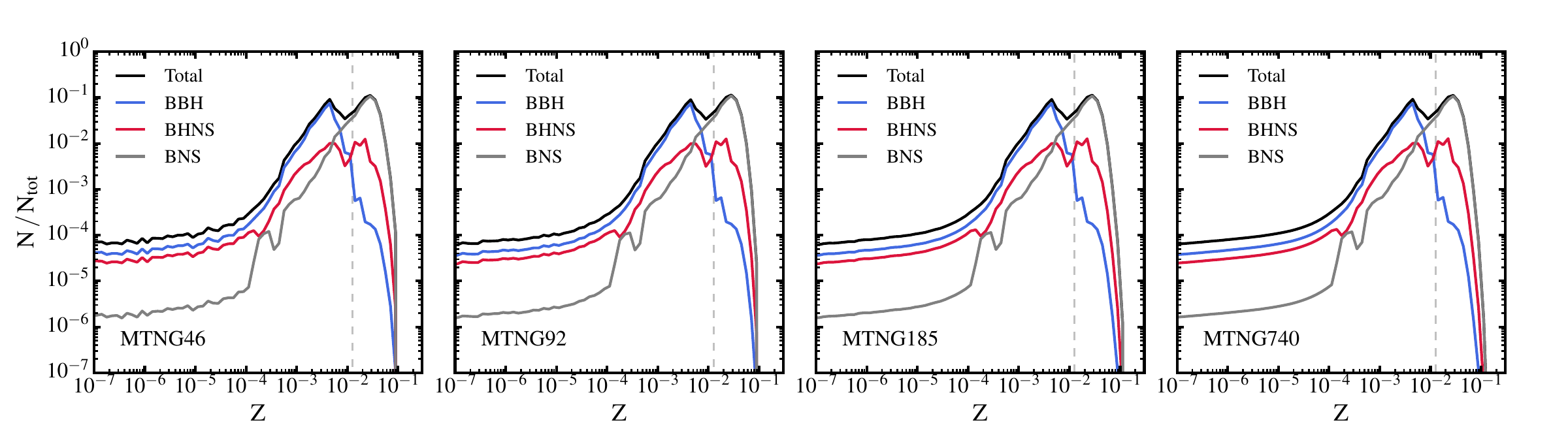}
      \caption{Distribution of GW progenitor events as a function of the metallicity of the parent stellar particle in four different boxes of the MTNG simulations: MTNG46, MTNG92, MTNG185, and MTNG740 (from left to right). The metallicity distribution of the total merger events, along with the contributions from different compact binary merger channels, is displayed as indicated in the legend. All distributions are normalized to the total number of merger events. As in Fig.~\ref{fig6b}, the dashed vertical lines mark the value of $Z_\odot$ adopted in the simulations.
              }
         \label{fig8b}
   \end{figure*}

\begin{figure*}
        \centering
        \includegraphics[width=\textwidth]{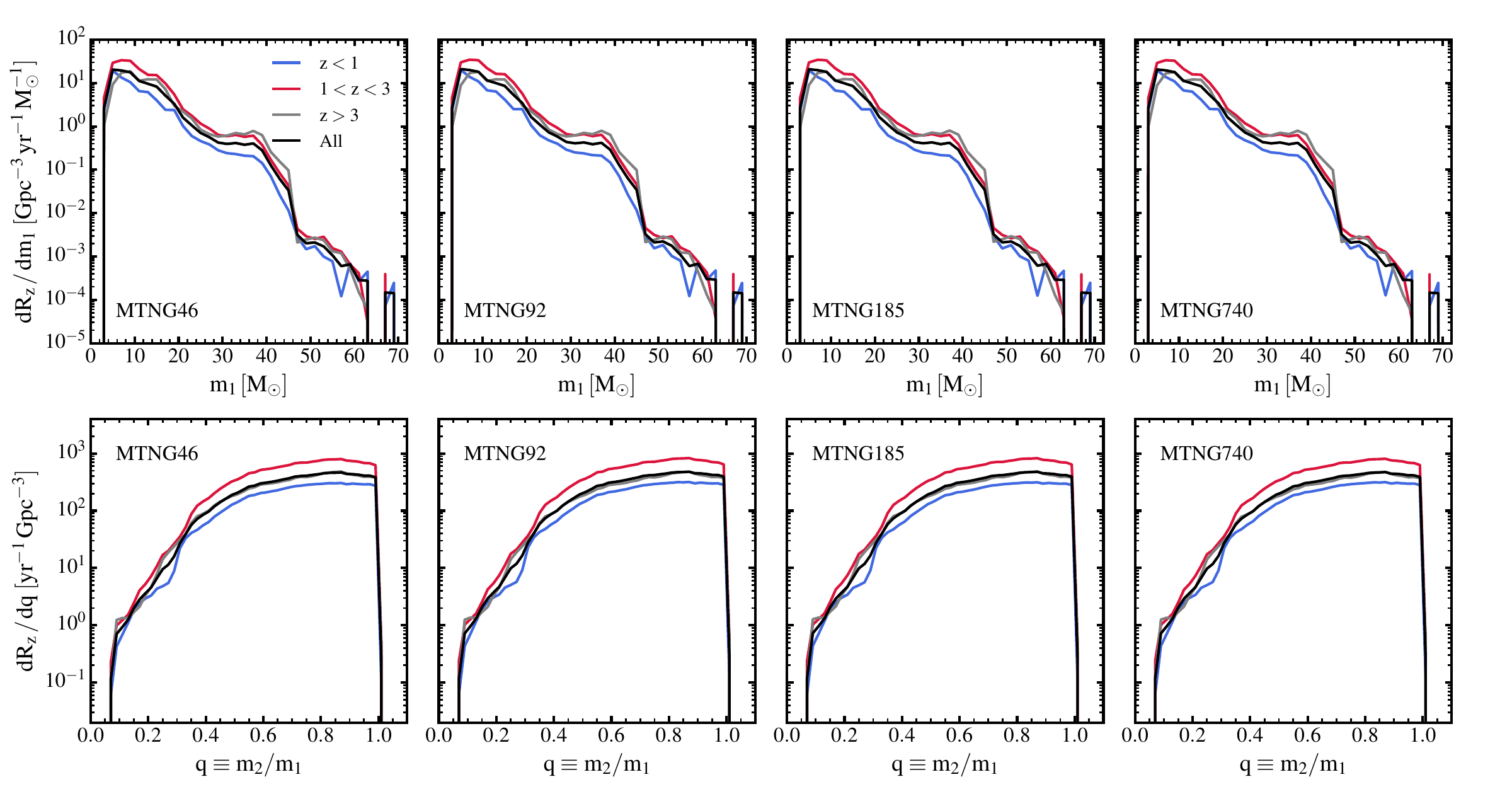}
        
      \caption{Redshift evolution of differential merger rates of BBH merger GW progenitor events in four different boxes of the MTNG simulations: MTNG46, MTNG92, MTNG185,  and MTNG740 (from left to right). In the top row the distributions are displayed as a function of the mass of the primary black hole ($m_1$) of the merging compact binary systems, whereas the bottom row shows the distributions in terms of the mass ratio $q \equiv m_2 / m_1$. Merger rates are normalized relative to the comiving volume of each simulation and to the time span (in Gyr) covered by the redshift intervals reported in the legend.}
         \label{fig9b}
    \end{figure*}

\begin{figure*}
        \centering
        \includegraphics[width=\textwidth]{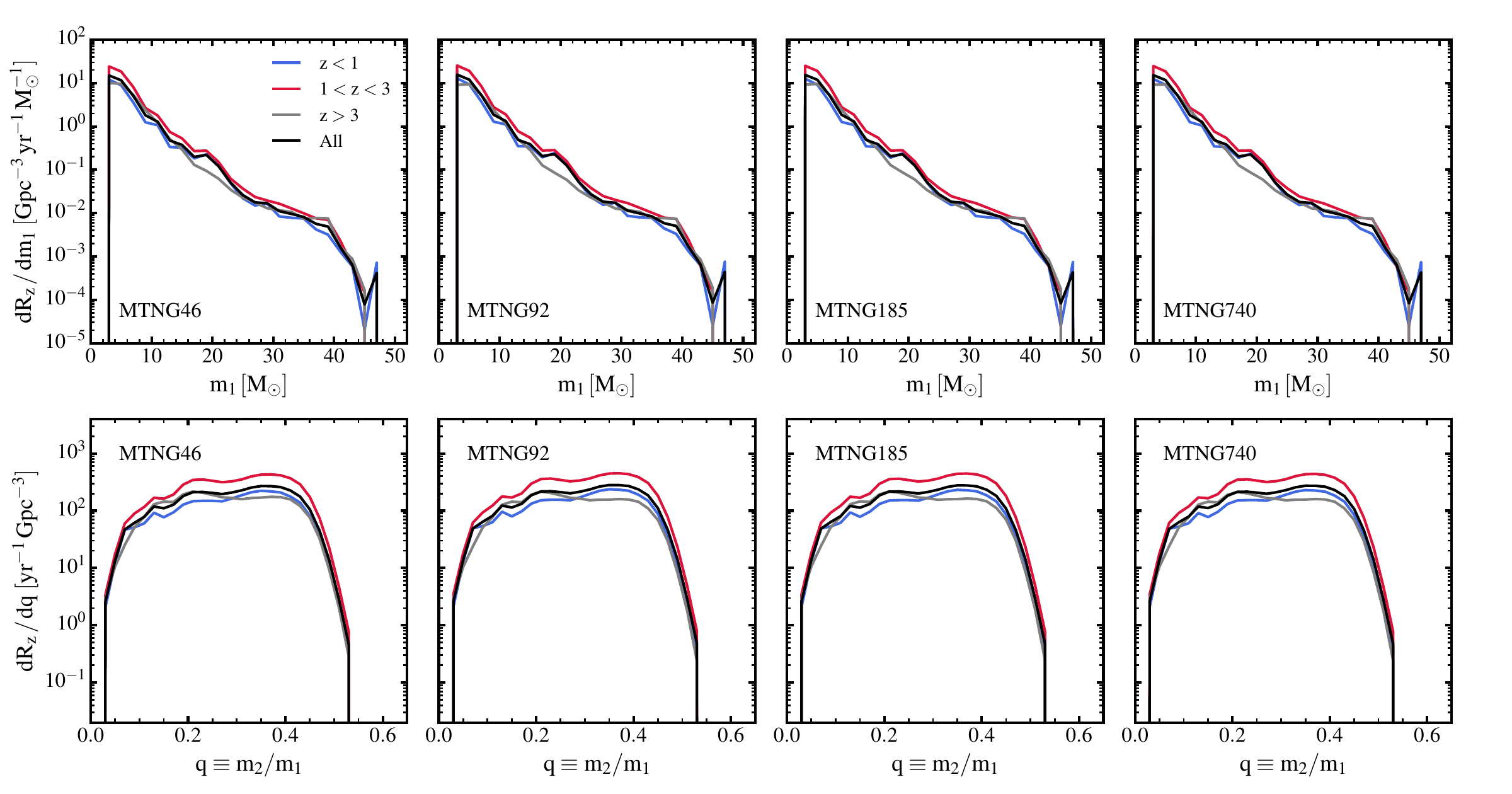}
        
      \caption{The same as Fig.~\ref{fig9b}, but for BHNS merger GW progenitor events.
             }
      \label{fig10b}
    \end{figure*}

\begin{figure*}
        \centering
        \includegraphics[width=\textwidth]{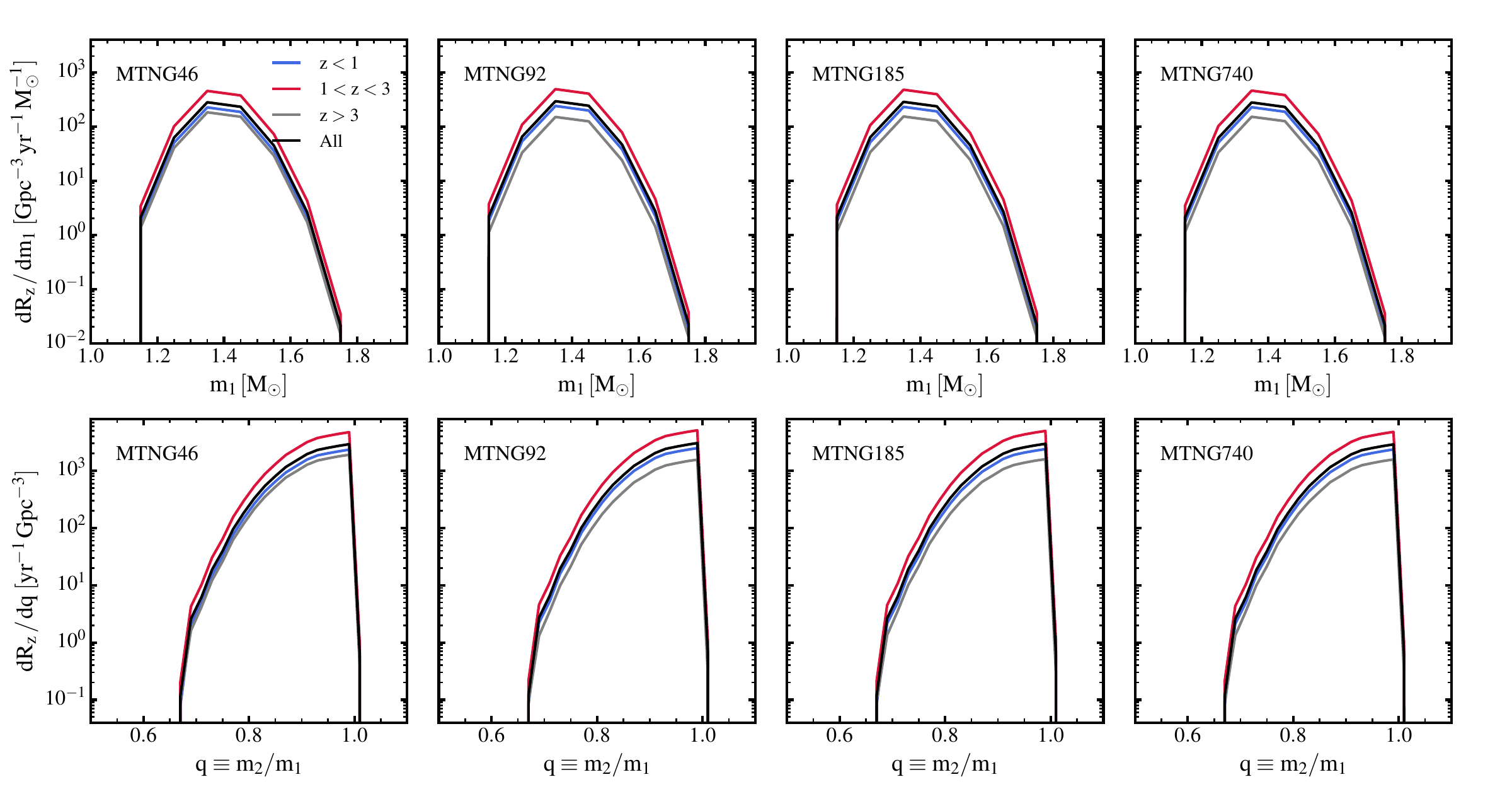}
        
      \caption{The same as Fig.~\ref{fig9b}, but for BNS merger GW progenitor events.
             }
    \label{fig11b}
    \end{figure*}

Each figure below is organized consistently. For each row, panels from left to the right  display  results for simulations with progressively increasing volume. We decided to also include MTNG740 results (already presented in the main body of the paper) as reference to facilitate comparison. The figures are displayed in the same order as discussed in the main text: comoving merger rates (Fig.~\ref{fig3b}), evolution of the binary merger efficiency per unit stellar mass (Fig.~\ref{fig4b}), delay time distribution (Fig.~\ref{fig5b}), binary merger efficiency as a function of metallicity (Fig.~\ref{fig6b}), two-dimensional histogram of SFRD as a function of lookback time and metallicity (Fig.~\ref{fig7b}), distribution of merger events as a function of the metallicity of the parent stellar particle (Fig.~\ref{fig8b}), and mass functions for BBH (Fig.~\ref{fig9b}), BHNS (Fig.~\ref{fig10b}), and BNS (Fig.~\ref{fig11b}, included here for completeness) mergers.

Owing to the constant mass resolution across all simulations, the results are largely consistent among the different volumes. The only noticeable difference arises in the smaller simulation volumes, where sampling noise becomes more evident because fewer GW progenitor events are generated. This effect is most pronounced in the MTNG46 box, which represents a relatively small cosmological volume, but it becomes progressively less significant in larger boxes. Therefore, relatively modest simulated volume -- which are less computationally expensive to run, post-process and analyze -- can be used to calibrate \arepogw\ parameters and test alternative version of merger tables produced by binary population synthesis simulations. This approach is extremely helpful for assessing the performance of different GW progenitor models, which can be employed, for instance, to address the tension with LVK results regarding the expected rate of BBH mergers.

\end{appendix}

\end{document}